\documentclass[%
  reprint,
  superscriptaddress,
  nofootinbib,
  longbibliography,
  amsmath,amssymb,
  aps,
  prd
  ]{revtex4-2}

\usepackage{graphicx}

\graphicspath{{./}{figures/}} 

\usepackage{dcolumn}
\usepackage{bm}
\usepackage{hyperref}
\hypersetup{
  colorlinks=true,        
  linkcolor=blue,         
  citecolor=cyan,         
}
\usepackage{colordvi}
\usepackage[dvipsnames]{xcolor}
\usepackage{units}
\usepackage{mathtools}
\usepackage{xspace}
\usepackage{acronym}
\usepackage{tabularx}
\usepackage{booktabs}
\usepackage[utf8]{inputenc}
\usepackage{makecell}
\usepackage{wrapfig}
\usepackage[normalem]{ulem}

\newcommand{\ra}{\rightarrow}
\newcommand{\tn}[1]{\textnormal{#1}}
\newcommand{\dd}{\tn{d}}

\newcommand{\secref}[1]{Sec.~\ref{sec:#1}}
\newcommand{\appref}[1]{Appendix~\ref{sec:#1}}
\newcommand{\figref}[1]{Fig.~\ref{fig:#1}}
\newcommand{\tabref}[1]{Table~\ref{tab:#1}}
\renewcommand{\eqref}[1]{Eq.~(\ref{eq:#1})}

\newcommand{\ie}[0]{i.e.\@\xspace}
\newcommand{\eg}[0]{e.g.\@\xspace}
\newcommand{\cf}[0]{cf.\@\xspace}

\newcommand{\bnsnurates}[0]{\textsc{bns\_nurates}\@\xspace}

\newcommand{\nue}[0]{\nu_{\tn{e}}}
\newcommand{\anue}[0]{\bar{\nu}_{\tn{e}}}
\newcommand{\nux}[0]{\nu_{\tn{x}}}
\newcommand{\anux}[0]{\bar{\nu}_{\tn{x}}}
\newcommand{\ye}[0]{Y_{\tn{e}}}
\newcommand{\p}[0]{\tn{p}}
\newcommand{\n}[0]{\tn{n}}
\renewcommand{\em}[0]{\tn{e}^{-}}
\newcommand{\ep}[0]{\tn{e}^{+}}

\usepackage{xcolor}
\usepackage{amsmath}
\usepackage{cancel}
\usepackage{amssymb}
\usepackage{comment}
\usepackage[framemethod=TikZ]{mdframed}

\newcommand{\mone}{\texttt{M1} }

\newacro{NSE}{nuclear statistical equilibrium}
\newacro{GCE}{galactic chemical evolution}
\newacro{EOS}{equation of state}
\newacroplural{EOS}{equations of state}
\newacro{BNS}{binary neutron star}
\newacro{NS}{neutron star}
\newacro{BH}{black hole}
\newacro{CC}{charged-current}
\newacro{CCSN}{core-collapse supernova}
\newacroplural{CCSN}[CCSNe]{core-collapse supernovae}
\newacro{GW}{gravitational wave}
\newacro{DOF}{degree of freedom}
\newacroplural{DOF}{degrees of freedom}
\newacro{NR}{numerical relativity}
\newacro{LK}{leakage}
\newacro{LR}{low resolution}
\newacro{SR}{standard resolution}
\newacro{HR}{high resolution}
\newacro{RMF}{relativistic mean field}
\newacro{EM}{electromagnetic}
\newacro{AMR}{adaptive mesh refinement}
\newacro{ISM}{interstellar medium}
\newacro{SN}{supernova}
\newacro{Mya}{Myr ago}
\newacro{OPE}{one-pion exchange}
\newacro{MHD}{magnetohydrodynamics}
\newacro{GRMHD}{general-relativistic magnetohydrodynamics}
\newacro{GRLES}{general-relativistic large-eddy simulation}
\newacro{NEPS}{neutrino-electron/positron scattering}
\newacro{LTE}{local thermodynamical equilibrium}
\newacro{QGP}{quark-gluon plasma}
\newacro{GPU}{graphical processing unit}
\newacro{CPU}{central processing unit}

\begin{document}


\title{Open-source library for performance-portable neutrino reaction rates:\\
Application to neutron star mergers}

\author{Leonardo Chiesa}
\email{leonardo.chiesa@unitn.it}
\affiliation{Dipartimento di Fisica, Università di Trento, via Sommarive 14,
  38123 Trento, Italy}
\affiliation{INFN-TIFPA, Trento Institute for Fundamental Physics and
  Applications, via Sommarive 14, 38123 Trento, Italy}
  
\author{Maitraya Bhattacharyya}
\affiliation{Institute for Gravitation \& the Cosmos, The Pennsylvania State
  University, University Park, Pennsylvania 16802, USA}
\affiliation{Department of Physics, The Pennsylvania State
  University, University Park, Pennsylvania 16802, USA}
  
\author{Filippo Mazzini}
\author{Federico Maria Guercilena}
\author{Albino Perego}
\affiliation{Dipartimento di Fisica, Università di Trento, via Sommarive 14,
  38123 Trento, Italy}
\affiliation{INFN-TIFPA, Trento Institute for Fundamental Physics and
  Applications, via Sommarive 14, 38123 Trento, Italy}

\author{David Radice}
\affiliation{Institute for Gravitation \& the Cosmos, The Pennsylvania State
  University, University Park, Pennsylvania 16802, USA}
\affiliation{Department of Astronomy \& Astrophysics, The Pennsylvania State
  University, University Park, Pennsylvania 16802, USA}


\begin{abstract}
A realistic and detailed description of neutrinos in \ac{BNS} mergers is essential to build reliable
models of such systems. To this end, we present
\bnsnurates, a novel open-source numerical library designed for the efficient on-the-fly computation of neutrino
interactions, with particular focus on regimes relevant to \ac{BNS} mergers. \bnsnurates targets a higher level of accuracy and realism in the implementation of commonly
employed reactions by accounting for relevant 
microphysics effects on
the interactions, such as weak magnetism and mean field effects.
It also includes the contributions of inelastic neutrino scattering off electrons and positrons
and (inverse) nucleon decays.
Finally,
it offers a way to reconstruct the neutrino distribution function in the framework of moment-based transport schemes.
As a first application, 
we compute both energy-dependent and energy-integrated neutrino emissivities and opacities for
conditions extracted from a \ac{BNS} merger simulation with \mone transport scheme.
We find some qualitative differences in the results when considering the impact of the additional relevant
reactions and of microphysics effects. For example,
neutrino-electron/positron scattering reactions are important for the energy exchange of heavy-type neutrinos as they do not undergo semileptonic
charged-current processes, when $\mu^\pm$ are not accounted for. Moreover, weak magnetism and mean field effects can significantly modify the
contribution of $\beta$ processes for electron-type (anti)neutrinos, increasing at the same time
the importance of (inverse) neutron decays.
The improved treatment for the reaction rates also modifies the conditions at which neutrinos decouple from matter in the system, potentially affecting their emission spectra.
\end{abstract}


\maketitle

\acresetall

\section{Introduction}
\label{sec:introduction}

The inclusion of detailed and realistic microphysics input in the description of
\ac{BNS} mergers is crucial to construct reliable models that can be
compared with astrophysical observations. Among the different aspects to be
considered, the interplay between neutrinos and nuclear matter must be modeled
accurately as it may generate some distinct fingerprints on the dynamics of the
coalescence (see, for example, \cite{Foucart:2022bth} for a recent review). In particular, neutrinos transport and redistribute energy and
momentum while diffusing across the system, see, \eg, \cite{Eichler:1989ve,Ruffert:1996by,Rosswog:2003rv,Dessart:2008zd,Perego:2014fma}. As a consequence,
they can affect the properties and the stability of the remnant, see, \eg, \cite{Loffredo:2022prq,Gieg:2024jxs,Pajkos:2024iml},
possibly influencing the emission of gravitational waves
in the merger aftermath \cite{Foucart:2015gaa,Zappa:2022rpd,Most:2022yhe}.
In the high-density core of a noncollapsed remnant, they may
also form a trapped and degenerate gas that contributes in a nontrivial way to
the total pressure support of the system \cite{Foucart:2015gaa,Perego:2019adq,Zappa:2022rpd,
Espino:2023dei,Pajkos:2024iml}. Neutrino interactions are also a critical input as they set the
neutron-to-proton content of ejected matter, with direct consequences on the
nucleosynthesis yields and on the associated kilonova signal \cite{Wanajo:2014wha,Metzger:2014ila,Martin:2015hxa,Foucart:2015gaa,Sekiguchi:2016bjd,Fujibayashi:2017xsz,Radice:2018pdn,Fujibayashi:2020qda,Just:2021cls,Just:2021vzy,Radice:2023zlw}. Recent
studies have highlighted how the ejecta properties and in particular the degree of leptonization depend on the neutrino transport and on the neutrino-matter interactions, see, \eg, \cite{Foucart:2018gis,Foucart:2024npn,Espino:2023mda,Cheong:2024buu}. This can, for example, affect the
production of light elements by \ac{BNS} coalescences, which may have been
historically underestimated as a result of the approximated neutrino treatment, see, \eg,
\cite{Perego:2020evn,Domoto:2021xfq,Chiesa:2023jno}. 
Other aspects concerning the role
of neutrinos in the context of \ac{BNS} mergers include the possible involvement
in the formation of relativistic jets \cite{Eichler:1989ve,Rosswog:2003tn,Dessart:2008zd,Zalamea:2010ax,Perego:2017fho,Musolino:2024sju} and the
occurrence of flavor-changing oscillations on time scales and length scales relevant for
the dynamics and for the matter ejection, see, \eg, \cite{Zhu:2016mwa,Frensel:2016fge,Wu:2017drk,George:2020veu,Nagakura:2023wbf,Grohs:2023pgq,Froustey:2023skf}. Finally, the characterization and possible detection of
neutrino emission from \ac{BNS} mergers is also an item of interest in its own
right \cite{Cusinato:2021zin}.

Neutrinos are numerically evolved in \ac{BNS} merger simulations by
approximately solving the Boltzmann transport equation. 
First attempts to include neutrino physics in \ac{BNS} mergers relied on leakage schemes or on hybrid schemes that coupled leakage prescriptions for the optically thick regime and ray tracing or moment schemes for the optically thin regime, see, \eg, \cite{Ruffert:1996by,Rosswog:2003rv,Perego:2014fma,Sekiguchi:2015dma,Sekiguchi:2016bjd,Radice:2018pdn,Ardevol-Pulpillo:2018btx}.
Recently, there have been
some efforts aimed at improving the precision and reliability of employed
transport schemes in \ac{BNS} mergers by means of two-moment schemes
\cite{Foucart:2015gaa,Foucart:2015vpa, Foucart:2016rxm,Radice:2021jtw, Schianchi:2023uky,Musolino:2023pao, Cheong:2023fgh}
or Monte Carlo based ones \cite{Miller:2019dpt, Miller:2019gig, Foucart:2020qjb, Foucart:2021mcb, Foucart:2022kon}.
Along with a robust transport scheme, feeding
accurate neutrino rates into the source term of the transport equation is of
paramount importance for modeling the correct neutrino dynamics, particularly in
the regions where neutrinos transition from equilibrium conditions to
free-streaming propagation. Dedicated software libraries have been designed for
such a purpose, \eg, \cite{OConnor:2014sgn, Ng:2023syk}, often resting upon the
expertise that has accumulated in the modeling of \acp{CCSN}. 
Recent studies have highlighted the importance of detailed microphysics to predict neutrino emissivities and absorptivities in hot and dense matter with high accuracy \cite{Bacca:2011qd,Roberts:2012um,Martinez-Pinedo:2012eaj,Roberts:2016mwj,Fischer:2018kdt,Guo:2020tgx,Oertel:2020pcg,Guo:2024opk}.
Possible relevant effects for the rate calculations of both neutral- and charged-current processes include full kinematics treatment, weak magnetism and pseudoscalar terms, energy-dependent nucleon form factors, and mean field energy shifts due to nuclear interactions.
Another essential aspect is the inclusion of all of the relevant degrees of freedom, together with the most important associated reactions. In the context of \ac{BNS} mergers, leptonic and semileptonic reactions involving muons should be also considered, see, \eg,
\cite{Fischer:2018kdt,Guo:2020tgx,Guo:2024opk,Gieg:2024jxs,Pajkos:2024iml,Ng:2024zve}.
Despite these progresses, the characterization of neutrino interactions in neutron star mergers is still partial. Because of both an incomplete knowledge of the reaction rates and limitations in our computational capability, the description of the various reactions that are typically included
in \ac{BNS} merger modeling is often oversimplified, \eg, by adopting crude approximations
on the matrix elements, by ignoring the impact of the
surrounding nuclear medium, by assuming neutrinos in
\ac{LTE} conditions in any regime, or by relying on precomputed and tabulated rates,
therefore introducing additional errors associated with interpolation procedures.
Furthermore, the set of reactions that are considered is often not exhaustive,
as there are processes whose impact in \ac{BNS} merger conditions could be
significant, but have not been explored yet in detail. As an example, the inelastic
scattering of neutrinos off electrons and positrons (proven to be relevant for
the dynamics of \acp{CCSN} and not usually included in \ac{BNS} merger
simulations), has been found by recent studies to be possibly
impactful in the merger aftermath \cite{Cheong:2024cnb}.

To tackle some of these issues, we have developed and present \bnsnurates, a novel
open-source numerical library designed for the computation of neutrino-matter interaction
rates in the regimes relevant for \ac{BNS} mergers. This library provides an
improved treatment of neutrino interactions, targeting a higher level of
accuracy and realism compared to commonly employed neutrino rates prescriptions.
\bnsnurates includes reactions that have not been extensively investigated in the
context of \ac{BNS} mergers, such as the inelastic neutrino-electron and
neutrino-positron scattering and (inverse) nucleon decays. It also accounts
for the impact of microphysics effects on some of the included interactions,
such as recoil, phase-space and weak magnetism effects for scattering off nucleons,
and $\beta$ processes and in-medium effects for $\beta$ processes.
\bnsnurates is able to compute both spectral (energy-dependent) and gray
(energy-integrated) neutrino emissivities and opacities for given thermodynamic conditions, by
prescribing a functional form for the distribution function of neutrinos. The
library goes beyond the assumption of \ac{LTE} conditions, as it
reconstructs the neutrino distribution function starting from the information
encoded in the first gray radiation moments, assuming a linear combination
between the contribution of trapped and free-streaming neutrinos. Despite its
focus on physical realism, the library has been designed to be computationally
efficient in order to be coupled to transport schemes for the on-the-fly evaluation of
neutrino opacities and emissivities, avoiding table interpolations. The computation is
additionally accelerated thanks to the integration with the \textsc{Kokkos}
library, which allows to offload the evaluation at different grid points onto
parallel \acp{GPU}. \bnsnurates is publicly available \cite{Perego_BNS_NURATES_2025}
and will be actively
maintained with the aim of further improving its accuracy, completeness and
computational performance.

In this work we first discuss the physical content of \bnsnurates in terms of the 
neutrino-matter interactions and of the level of accuracy that it implements.
We then apply \bnsnurates to the evaluation of both energy-dependent and
energy-integrated neutrino emissivities and opacities for a set of thermodynamic conditions
representative of the postmerger phase of a \ac{BNS} system, extracted from a
simulation employing a gray \mone neutrino transport scheme, in a
postprocessing analysis. We discuss in detail what is the contribution of the
different reactions and of microphysics effects in such conditions. Our goal is, on the one hand,
to assess which processes are most relevant, and, on the other hand, to assess if the
improved treatment could be impactful for the dynamics of neutrinos in the
system.

We find that reactions and microphysics effects that are not always included
in sophisticated simulations, but have been implemented into the library, have a significant
impact on the total neutrino rates. In particular, we show that the inclusion of
neutrino scattering off electrons and positrons provides an additional
contribution to the opacity that is sizable wherever charged-current processes
are not relevant, \ie, deep inside the remnant in the case of electron
(anti)neutrinos and in general for heavy-type (anti)neutrinos. For the latter,
we find that the emission spectrum is modified as their neutrino surfaces are
pushed out to larger radii. Also mean field and weak magnetism effects prove to be
influential, depending on the conditions. Mean field effects enhance both the emissivity
and opacity of $\nue$'s and $\anue$'s when the density is large enough,
\ie, for rest mass density $\rho \gtrsim \unit[10^{13}]{g \, cm^{-3}}$. The weak magnetism effect
instead makes electron antineutrino's interactions less frequent, in particular
when they have large energies, and remains relevant in regions at lower
densities as well, including the ones where $\anue$ typically decouples from
matter. 

The paper is structured as follows: in \secref{methods} we present a general
overview about the formalism of the collision integral for neutrino transport,
as well as the physical content of the library in terms of reactions and
microphysics effects implemented and
of the strategy followed to reconstruct the neutrino distribution functions.
In \secref{postproc_strategy}, we elaborate on how we extract representative
thermodynamic conditions from a \ac{BNS} merger simulation to evaluate
neutrino reactions.
Then, in Secs.~\ref{sec:spectral} and \ref{sec:gray} we discuss the
results obtained for neutrino emissivities, opacities, and surfaces in the postprocessing
analysis, separately for the energy-dependent and energy-integrated cases. We also discuss
the numerical performance of \bnsnurates in \secref{performance}.
Finally, we summarize the work and draw our conclusions in \secref{conclusions}.
Appendixes \ref{sec:reactions_implementation} to \ref{sec:additional_plots} are dedicated to
the elucidation of technical details and the presentation of additional results.

\section{Methods}
\label{sec:methods}

This section summarizes and elaborates on the formalism and equations underlying
the framework in which the \bnsnurates library has been developed.
In the following, Greek (italic) letters run over spacetime (space) indexes.
However, the notation $A_x$ for some generic quantity $A$ indicates dependence on
the neutrino species $\left(x\in\left\{\nue,\anue,\nux,\anux\right\}\right)$,
and any ambiguity is resolved in the text. We use $\nux$ ($\anux$) as a generic
notation for indicating either $\nu_\mu$ ($\bar{\nu}_\mu$) or $\nu_\tau$ ($\bar{\nu}_\tau$)
neutrinos without distinction, as there are for now no interactions in our library
yielding different rates for these two flavors. In some expressions, the symbol $\nu$
stands for a neutrino of any species. We also adopt the convention to set the Boltzmann constant $k_B=1$, so that temperatures are expressed in units of
energy throughout.

\subsection{Collision integral and stimulated absorption}
\label{sec:boltzmann_equation}

The transport of neutrinos is regulated by the interactions with the background
matter field (hereafter fluid) and classically described by the Boltzmann equation
for the neutrino distribution functions $f_x(x^\mu,p^\mu)$\footnote{In this section, several quantities such as $p^\mu$ pertain to a given neutrino species, \ie, $p^\mu$ should be denoted as $p_x^\mu$. We employ the former notation in the interest of readability.}:
\begin{equation}
  p^\alpha \left[ \frac{\partial f_x}{\partial x^\alpha}  
    + \Gamma_{\alpha \gamma}^i p^\gamma \frac{\partial f_x}{\partial p^i}
  \right] = (-p^\mu u_\mu ) \, B_x (x^\mu, p^\mu,f_x) \, ,
  \label{eq:spectral_boltzmann_eq}
\end{equation}
where $x^\mu$, $u^\mu$, and $p^\mu$ are the spacetime coordinates, the fluid four-velocity, and the neutrino
four-momentum in the laboratory frame, respectively.
The four-momentum satisfies $p^\mu p_\mu = 0$, \ie, we consider massless neutrinos. In
\eqref{spectral_boltzmann_eq}, $\Gamma^i_{\alpha\gamma}$ are the Christoffel symbols of the
background spacetime. We can decompose the four-momentum $p^\alpha$ as
\begin{equation}
    p^\alpha = k(u^\alpha + l^\alpha) \, ,
    \label{eq:p_alpha}
\end{equation}
where $k = - p^\mu x_\mu$ is the neutrino energy measured in the fluid frame
and $l^\alpha$ is a unit four-vector orthogonal to
$u^\alpha$, encoding the angular dependence of the neutrino momenta (note that
$l_\alpha l^\alpha=l^2=1$ and $u_\alpha l^\alpha=0$). The angular dependence can
also be expressed in terms of the angular variables $\Omega=\{\theta,\phi\}$,
defined by the direction of propagation of neutrinos in the frame comoving with the fluid. In the following, we will omit
the dependence on $x^\mu$ for simplicity and specify the one on $p^\mu$ in terms of ($k,\Omega$).
On the right-hand side of
\eqref{spectral_boltzmann_eq}, $B_x \equiv \left(\dd f_x/ \dd\tau
\right)_{\tn{coll}}$ is the collision integral (defined in terms of the fluid
proper time, $\tau$) which accounts for the neutrino interactions
\footnote{Hereafter, it is understood that integrals over the energy
range from 0 to $+\infty$, while integrals over angular dependence span the
whole two-sphere.}:
\begin{widetext}
\begin{equation}
  \begin{split}
    B_x(k,\Omega,f_x) &= \\
       &\left(1 - f_x\right)
         j_{\beta,x}\left(k \right)
         - f_x \, c\,\lambda^{-1}_{\beta,x}\left(k \right)  \\
        +& \left(1 - f_x\right)
        \, \int \dd k' \frac{k'^2}{(hc)^3} \int \dd\Omega'
        \left(1-\Bar{f}'_x \right)
        R^{{\tn{pro}}}_x\left(k, k', \omega\right) -
        f_x \int \dd k' \frac{k'^2}{(hc)^3} \int \dd\Omega' \Bar{f}'_x
        R^{{\tn{ann}}}_x\left(k, k', \omega\right)  \\
        +& \left(1 - f_x\right)
        \int \dd k' \frac{k'^2}{(hc)^3} \int \dd \Omega'
        f'_x R^{{\tn{in}}}_x\left(k, k', \omega\right) -
        f_x \int \dd k' \frac{k'^2}{(hc)^3} \int \dd \Omega'
        \left( 1 - f'_x \right)
        R^{{\tn{out}}}_x\left(k, k', \omega\right) \, ,
  \end{split}
  \label{eq:collision_term_extended}
\end{equation}
\end{widetext}
where $\Bar{f}_x=\Bar{f}_x(k,\Omega)$ is the distribution function of the antiparticle
relative to the (anti)neutrino of flavor $x$. The primed notation for $f'$ and $\Bar{f}'$
indicates a dependence on the integration variables $(k',\Omega')$. By inspecting the
different terms that appear on the right-hand side of \eqref{collision_term_extended},
we can identify the contribution to the collision integral of different classes of reactions.
The first row accounts for semileptonic charged-current processes, hereafter $\beta$ processes,
where a single neutrino takes part to the interaction.
The second row describes reactions involving
the emission or absorption of $\nu\Bar{\nu}$ pairs, quantified by the production and
annihilation kernels, $R^{\tn{pro}}_x$ and $R^{\tn{ann}}_x$, summed over the different
interactions which fall within this class.
Finally, the third row indicates the contribution of neutrino scattering reactions
in terms of the $R^{\tn{in}}_x$ and $R^{\tn{out}}_x$ kernels.
The kernels depend on the energies of both (anti)neutrinos taking part in the interaction
and on the angle between their propagation directions, which we denote as $\omega$.
The latter can be expressed in terms of the angular variables of the individual particles as
\begin{equation}
  \omega = \cos\theta\cos\theta' + \sin\theta\sin\theta' \cos(\phi-\phi') \, .
\end{equation}

We rewrite \eqref{collision_term_extended} by grouping together the positive
and contributions of all reactions, including the
scattering processes:
\begin{equation}
  \begin{split}
    B_x&(k,\Omega,f_x) = \left(1 - f_x\left(k,\Omega\right)\right)
         \left\{\sum_l j_{l,x}\left(k, \Omega\right)\right\}\\
       &- f_x\left(k,\Omega\right)
         \left\{\sum_l c\,\lambda^{-1}_{l,x}\left(k, \Omega\right)
         \right\} \, ,
  \end{split}
  \label{eq:collision_term}
\end{equation}
where the index $l$ runs over all reactions. 
This notation makes explicit how each 
process contributes to the production and removal of neutrinos and
is chosen to facilitate the comparison between the different reactions.
In the first term, the spectral emissivity, $j_{l,x}$, once
integrated over the phase space, provides the number of neutrinos
of flavor $x$ and four-momentum $(k,\Omega)$ that are produced per unit time by
the $l$th reaction. This is
weighted by the Pauli blocking factor of the final state, $(1-f_x)$. Similarly,
the spectral absorptivity $c \lambda_{l,x}^{-1}$ (or inverse mean free path
$\lambda_{l,x}^{-1}$) quantifies the removal of neutrinos of given species and
momentum per unit time, and is weighted by the neutrino occupation number $f_x$.
We do not report explicitly the expression of $j_{l,x}$ and $c\lambda_{l,x}^{-1}$
for each process
as they can be easily identified in \eqref{collision_term_extended} as the terms that multiply
$(1-f_x)$ and $f_x$, respectively, depending on the reaction class.

Finally, \eqref{collision_term} can also be rewritten in the following way:
\begin{equation}
  \begin{split}
    B_x&(k,\Omega,f_x) = \sum_l j_{l,x}(k,\Omega) \\
       &- f_x(k,\Omega)\sum_l K_{l,x}(k,\Omega) \, ,
  \end{split}
  \label{eq:collision_term_stim_abs}
\end{equation}
where $K_{l,x} \equiv j_{l,x} + c\,\lambda^{-1}_{l,x}$ is the stimulated
(spectral) absorptivity. This formalism will come in handy when discussing the
source term of a gray \mone transport scheme in \secref{energy_integrated_rates}.
The spectral inverse mean free paths presented in \secref{spectral} follow the notation appearing
in \eqref{collision_term}.

\subsection{Neutrino reactions}
\label{sec:neutrino_reactions}

\begin{table*}[t]
  \begin{center}
    \begin{tabularx}{\textwidth}{ >{\centering\arraybackslash}X  >{\centering\arraybackslash}X}
      \toprule
      $\beta$ processes
      \cite{Bruenn:1985en,Oertel:2020pcg,Horowitz:2001xf,Hempel:2014ssa} &
      \makecell{
      $\p + \em \leftrightarrow \n + \nue\quad\n + \ep \leftrightarrow \p + \anue$\\
      $\p  \leftrightarrow \ep + \n + \nue\quad\n \leftrightarrow \em + \p + \anue$
      }\\[1pt]
      Electron-positron annihilation \cite{Bruenn:1985en,Pons:1998st} &
      $ \em + \ep \leftrightarrow \nu + \bar{\nu}$\\[1pt]
      Nucleon-nucleon bremsstrahlung \cite{Hannestad:1997gc,Fischer:2016boc} &
      $\tn{N} + \tn{N} \leftrightarrow \tn{N} + \tn{N} + \nu + \bar{\nu}$\\[1pt]
      Isoenergetic scattering off nucleons
      \cite{Bruenn:1985en,Horowitz:2001xf} &
      $\tn{N} + \nu \ra \tn{N} + \nu$\\[1pt]
      Inelastic scattering off electrons and positrons
      \cite{Bruenn:1985en,Mezzacappa:1993mab} &
      $\tn{e}^\pm + \nu \ra \tn{e}^\pm + \nu$\\
      \bottomrule
    \end{tabularx}
  \end{center}
  \caption{Reactions included in \bnsnurates}
  \label{tab:reactions}
\end{table*}

The actual form of $j_{l,x}$ and $\lambda^{-1}_{l,x}$ depends on the
neutrino-matter reaction under consideration. The set of reactions that are
currently included in \bnsnurates is summarized in \tabref{reactions}.
We distinguish between interactions involving a single neutrino and those
involving two (anti)neutrinos among the initial and final states. In the former case,
spectral emissivities and absorptivities are given by closed-form expressions, as one can integrate over the
phase spaces of all the other particles involved by assuming them to be in \ac{LTE} conditions. $\beta$ processes fall within this category.
In the second case, we describe neutrino interactions in terms of the corresponding
reaction kernels since the spectral emissivities and absorptivities
specifically depend on the
distribution function prescribed for the other (anti)neutrino participating
in the reaction. This category includes pair and scattering processes. For such interactions, we expand the reaction kernels in a Legendre series, such that
\begin{equation}
    R^{\genfrac(){0pt}{3}{\tn{pro}}{\tn{in}}}_{l,x}(k,k',\omega) = \sum_{m=0}^{+\infty}R^{\genfrac(){0pt}{3}{\tn{pro}}{\tn{in}},m}_{l,x}(k,k')
    \, P_m(\omega) \, ,
    \label{eq:kernel_leg_expansion_general}
\end{equation}
where $P_m(\omega)$ is the Legendre polynomial of order $m$ and
$R^{\genfrac(){0pt}{3}{\tn{pro}}{\tn{in}},m}_{l,x}$ is the $m$th coefficient
of the expansion. For the evaluation of spectral emissivities and absorptivities, we truncate the expansion at
$m = 0$.  As a consequence, $j_{l,x}$ and $\lambda_{l,x}^{-1}$ do not have any
angular dependence in the current version of \bnsnurates.
The implementation of angular dependent interactions is left as a future enhancement of the library.

Hereafter we briefly discuss the implementation of the various reactions included in \bnsnurates, while we leave the
presentation of the explicit formulas to \appref{reactions_implementation}.
\paragraph{$\beta$ processes.}
$\beta$ processes are interactions in which a single
neutrino induces a $n \leftrightarrow p$ conversion by coupling with the corresponding
lepton. They include electron and positron captures on nucleons, nucleon decays and the
corresponding inverse reactions.
We implement bare emissivities resulting from electron and positron captures
following Ref.~\cite{Bruenn:1985en} (see also \cite{Burrows:2004vq}), assuming
nonrelativistic nucleons and zero momentum transfer; while the ones of neutron and proton
decays are taken from Ref.~\cite{Oertel:2020pcg} and derived under the same
assumptions. To enhance the realism of the calculations, \bnsnurates implements
weak magnetism and in-medium effects for $\beta$ processes.
The weak magnetism effect is introduced to compensate for the fact that the
Pauli-tensor term of the nucleonic current is neglected when evaluating the matrix element
of the reaction. It is applied as an energy-dependent multiplicative factor to the bare
$j_{\beta,x}$, including also recoil and phase-space corrections, as discussed in Ref.~\cite{Horowitz:2001xf}.
For simplicity's sake, in the rest of the manuscript we will indicate the combination of these 
effects simply as weak magnetism.
In-medium (or mean field) effects instead account for the modification of bare rates induced by the interactions
between nucleons in the medium. The modification is stronger at higher densities, as the interaction
potential energy is on average larger. In a \ac{RMF} formalism, this effect is introduced
by replacing the bare nucleon mass with an effective one ($m_i \ra m_i^*$, $i=n,p$) and by
adding to the total single-particle energy the contribution of an interaction potential $U_i$.
Clearly, this is an \ac{EOS}-dependent effect.
The way in which the effective mass difference, $\Delta m^* \equiv m_n - m_p$, and the
interaction potential difference, $\Delta U \equiv U_n - U_p$,
modify the standard expression of $j_{\beta,x}$ has been 
discussed, \eg, in Refs.~\cite{Hempel:2014ssa,Oertel:2020pcg}.
The spectral absorptivity, $c \, \lambda_{\beta,x}^{-1}$, is obtained from the 
emissivity via detailed balance, \ie,
\begin{equation}
  c \, \lambda^{-1}_x(k) = j_x(k)\,e^{(k-\mu_x)/T} \,,
\end{equation}
where $T$ is the fluid temperature and $\mu_{\nue} = -\mu_{\anue} = \mu_\p -\mu_\n + \mu_{\tn{e}}$ is the
electron (anti)neutrino chemical potential at equilibrium, obtained from the relativistic chemical
potentials of neutrons ($\mu_\n$), protons ($\mu_\p$), and electrons ($\mu_{\tn{e}}$).

\paragraph{Pair processes.}
We account for the production and absorption of $\nu\bar{\nu}$ pairs via the
electron-positron annihilation and the nucleon-nucleon
bremsstrahlung (hereafter NN bremsstrahlung) and the corresponding inverse
reactions.
For each of the two interactions, the production and annihilation kernels are connected through the detailed
balance relation:
\begin{equation}
  R^{{\tn{ann}}}_{l,x}(k,k',\omega) = e^{(k+k')/T}\,R^{{\tn{pro}}}_{l,x}(k,k',\omega) \, .
\end{equation}
The kernel formulas for pair annihilation have been taken from
Refs.~\cite{Bruenn:1985en,Pons:1998st}, while the ones for NN bremsstrahlung from
Ref.~\cite{Hannestad:1997gc}, including the possibility for the latter to account for
in-medium modifications of the kernel according to Ref.~\cite{Fischer:2016boc}.

\paragraph{Scattering processes.}
We consider scattering reactions of neutrinos off both nucleons and $e^\pm$.
In this case as well, each of the \textit{in} and \textit{out}
scattering kernels are related through detailed balance:
\begin{equation}
  R^{{\tn{out}}}_{l,x}(k,k',\omega)=
  e^{(E_{\tn{t}}-E'_{\tn{t}})/T}\,R^{{\tn{in}}}_{l,x}(k,k',\omega)\,,
\end{equation}
where $\left(E_{\tn{t}}-E'_{\tn{t}}\right)$ is the energy transferred to the
target particle in the reaction. For conditions relevant to \ac{BNS} mergers, the scattering
off nucleons can be considered as quasielastic, \ie, the recoil is small due to the large mass
of the targets compared with the typical neutrino energies, so
that $R^{{\tn{in}}}_{\tn{iso},x}=R^{{\tn{out}}}_{\tn{iso},x} \equiv R_{\tn{iso},x}$.
The quasielastic (or isoenergetic) kernels follow Ref.~\cite{Bruenn:1985en} (or equivalently
\cite{Burrows:2004vq}) and account for phase-space, recoil and weak
magnetism effects for neutral-current reactions from Ref.~\cite{Horowitz:2001xf}.
In this case as well, we will refer to the combination of these effects simply as weak magnetism.
Note that for the evaluation of the isoenergetic spectral emissivities and absorptivities we expand the
kernel considering only $R^0_{\tn{iso},x}$, as mentioned before,
but we include also $R^1_{\tn{iso},x}$ when computing the energy-integrated scattering opacities
(see \secref{energy_integrated_rates}). In this latter case, before performing the energy integration,
it is also useful to apply the isoenergetic condition, \ie,
$R_{\tn{iso},x} = \tilde{R}_{\tn{iso},x}(k,\omega)\delta(k-k')$,
and write down its contribution to the collision integral as
\begin{equation}
    \begin{split}
      B^{\tn{iso}}_x(k) &= \frac{k^2}{(hc)^3}
               \int \dd \Omega' \times\\
             & \left[ f_x(k,\Omega') - f_x(k,\Omega) \right]
               \tilde{R}_{\tn{iso},x}(k, \omega) \, .
    \end{split}
    \label{eq:iso_collision_integral}
\end{equation}
For the same physical conditions, the scattering off $e^\pm$ is significantly inelastic and
its scattering kernels are modeled as in Refs.~\cite{Bruenn:1985en,Mezzacappa:1993mab}.
In the rest of the manuscript we will occasionally refer to the combination of $\nu \em$ and $\nu \ep$
scattering reactions as (inelastic) \ac{NEPS}.

\subsection{Energy-integrated emissivities and opacities}
\label{sec:energy_integrated_rates}

Following Ref.~\cite{Shibata:2011kx}, the energy-dependent source terms for the
radiation field equations in the moment-based transport formalism truncated to
second order can be written (in the local rest frame of the fluid) as
\begin{equation}
  S_{(k),x}^\alpha(k) =
  \frac{k^3}{(hc)^3}
  \int \dd\Omega \, B_x(k,\Omega,f_x)(u^\alpha+l^\alpha) \, .
  \label{eq:m1_spectral_source_term}
\end{equation}
The energy-integrated version of \eqref{m1_spectral_source_term} is defined in
Eq.~(5) of Ref.~\cite{Radice:2021jtw} and reads
\begin{equation}
  \begin{split}
    S^\alpha_x
    & \equiv
      \int \dd k\,S_{(k),x}^\alpha(k)  \\
    & = (\eta_x-\kappa'_{a,x}J_x) u^\alpha -
      (\kappa''_{a,x}+\kappa_{s,x})H_x^\alpha,
  \end{split}
  \label{eq:m1_gray_source_term}
\end{equation}
where $\eta_x$ is the gray neutrino emissivity, $\kappa'_{a,x}$ and
$\kappa''_{a,x}$ are the two gray absorption opacities and $\kappa_s$ is the
gray scattering opacity.
Note that $\kappa'_a$ and $\kappa''_a$ are in principle
different from each other.
$J_x$ and $H_x^\alpha$ are, respectively, the zeroth
and first gray radiation energy moments, namely the neutrino energy density and the
neutrino energy flux, \ie,
\begin{align}
  J_x &=
        \frac{1}{(hc)^3}
        \int \dd k ~ k^3
        \int \dd\Omega \, f_x(k,\Omega) \, ,
        \label{eq:m1_zeroth_energy_moment} \\
  H^\alpha_x &=
               \frac{1}{(hc)^3}
               \int \dd k ~ k^3
               \int \dd\Omega \, f_x(k,\Omega)\,l^\alpha \, .
               \label{eq:m1_first_energy_moment}
\end{align}

Often, gray transport schemes involve scalar equations for the evolution of
the neutrino number moments as well (see, \eg, \cite{Foucart:2016rxm,Andresen:2024mtt},
in which gray two-moment neutrino transports are employed in \ac{CCSN} simulations).
In this case, the zeroth-order moment is the neutrino number density,
$n_x$, defined as
\begin{equation}
  n_x =
  \frac{1}{(hc)^3}
  \int \dd k \, k^2
  \int \dd\Omega \, f_x(k,\Omega) \, .
  \label{eq:m1_zeroth_number_moment}
\end{equation}
These equations feature source terms of the form
\begin{equation}
    \Tilde{S}_x
    \equiv
    \int \dd k \, \Tilde{S}_{(k),x}(k) =
    \Tilde{\eta}_x-\Tilde{\kappa}_{a,x} n_x \, ,
    \label{eq:integratedsource_number}
\end{equation}
with corresponding emissivities and absorption opacities, $\Tilde{\eta}_x$ and
$\Tilde{\kappa}_{a,x}$. Equation~(\ref{eq:integratedsource_number}) represents the
energy-integrated version of
\begin{equation}
  \Tilde{S}_{(k),x}(k) =
  \frac{k^2}{(hc)^3}
  \int \dd\Omega \, B_x(k,\Omega) \, .
  \label{eq:spectralsourceterm_for_number}
\end{equation}

In order to evaluate the different coefficients in the moment expansion
of the source terms,
for each reaction we plug the corresponding collision integral
into Eqs.~(\ref{eq:m1_spectral_source_term}) and (\ref{eq:spectralsourceterm_for_number}), and 
we integrate over $k$. We then compare the
outcome of this procedure with Eqs.~(\ref{eq:m1_gray_source_term}) and (\ref{eq:integratedsource_number}),
respectively, and map the different terms according to the order of the neutrino moments.
In doing that, we find that the integrand defining $\kappa''_{a,x}$ contains the sum of
the stimulated spectral absorptivities. For the matter of coherence, when we map
the terms proportional to $u^\alpha$ in \eqref{m1_gray_source_term}, we define also $\kappa'_a$
in terms of $K_{l,x}$, with $\eta$ that is consequently specified in terms of the unblocked 
spectral emissivities.
In this way, it is possible to discuss the conditions under which $\kappa_a' = \kappa_a'' \equiv \kappa_a$.
In particular, this is realized when the angular integrations can be factored out from the ones over energy,
so that the former either become trivial or cancel out in the computation of the gray averages.
In order for this to happen, we truncate all the Legendre expansions of the inelastic kernels
at the monopole contribution (see \secref{neutrino_reactions}), so that spectral emissivities and
absorptivities do not depend
on angular variables. Furthermore, we consider neutrino distribution functions separable in their energy
and angular dependence, \ie, $f_x(k,\Omega) = g_x(k) \, h_x(\Omega)$, requiring that their angular part
satisfies the relation $\int \dd\Omega \, h_x(\Omega) = 4\pi$ (see \secref{nu_distribution}).
In this way, all the angular integrations can be carried out analytically, and gray emissivities
and absorption opacities can be obtained from the spectral ones 
in the following way:
\begin{align}
  \tilde{\eta}_x &=
                 \frac{4\pi}{(hc)^3} \int
                 \dd k \, k^{2} \, \sum_{l'} j_{l',x}(k) \, ,
                 \label{eq:gray_emissivity_number} \\
    \eta_x &=
                 \frac{4\pi}{(hc)^3} \int
                 \dd k \, k^{3} \, \sum_{l'} j_{l',x}(k) \, ,
                 \label{eq:gray_emissivity_energy} \\
  \tilde{\kappa}_{a,x} &=
                       \frac{4\pi}{c(hc)^3\,n_x}
                       \int \dd k
                       \, k^{2} \, g_x(k) \,
                       \sum_{l'} K_{l',x} \, ,
                       \label{eq:gray_abs_opacity_number} \\
  \kappa_{a,x} &=
                       \frac{4\pi}{c(hc)^3\,J_x}
                       \int \dd k
                       \, k^{3} \, g_x(k) \,
                       \sum_{l'} K_{l',x} \, ,
                       \label{eq:gray_abs_opacity_energy}
\end{align}
The sum over $l'$ runs over all inelastic reactions, \ie, all the reactions implemented
in the library except for the isoenergetic scattering off nucleons, which follows a different paradigm.
In fact, its contributions cancel out with each other at the lowest
order of the source terms' expansion, being an isoenergetic process.
As a consequence, it does not contribute to \eqref{integratedsource_number}.
In fact, it only enters in \eqref{m1_gray_source_term} with a coefficient, $\kappa_{s,x}$, which is
proportional to the neutrino energy flux.
Starting from the collision integral in \eqref{iso_collision_integral}, one obtains
\footnote{The simplified form of the collision integral
allows us to retain the first-order term of the Legendre
expansion of the kernel without introducing any nonanalytical
angular integration.}
\begin{equation}
  \begin{split}
    \kappa_{s,x} & = \frac{(4\pi)^2}{c(hc)^6 \, J_x}
                   \int \dd k \, k^5 g_x(k) \times \\
                   & \left[\tilde{R}_{\tn{iso},x}^0(k)-\frac{\tilde{R}_{\tn{iso},x}^1(k)}{3}\right] \, .
  \end{split}
  \label{eq:iso_scatteringcoefficient}
\end{equation}
Because of the different nature of the interaction, this 
coefficient is distinguished from $\kappa''_{a,x}$.
On the other hand, the cancellation does not occur in the case of
inelastic scattering reactions. Therefore, we consider the
scattering off $e^\pm$ as equivalent to a neutrino
emission/absorption process, contributing to $\eta_{x}$ ($\tilde{\eta}_x$) and $\kappa_{a,x}$
($\tilde{\kappa}_{a,x}$) together with the other inelastic reactions.
Notice that the evaluation of Eqs.~(\ref{eq:gray_emissivity_number})-(\ref{eq:iso_scatteringcoefficient})
requires us to perform a one-dimensional numerical integration over the neutrino energy in
the case of $\beta$ processes and isoenergetic scattering, and a two-dimensional
integration over the energies of the two (anti)neutrinos participating in the
interaction for the other classes
of reactions. The way in the numerical integrations are performed in the library is
detailed in \appref{quadrature_scheme}.

\subsection{Reconstruction of the neutrino distribution functions}
\label{sec:nu_distribution}

This section describes the procedure employed by \bnsnurates to reconstruct the neutrino
distribution function in the context of a gray transport scheme. 
We assume that the functional form of $f_x$ is separable, and we isolate its
energy and angular dependent parts, \ie, $f_x(k,\Omega)=g_x(k)\,h_x(\Omega)$. We
write the energy-dependent part $g_x$ as a linear combination of a trapped
contribution and a free-streaming one, labeled using the subscripts $_t$ and
$_f$, respectively:
\begin{equation}
  g_x(k) = w_{t,x} \, g_{t,x}(k) + w_{f,x} \, g_{f,x}(k) \,,
  \label{eq:g_nu_total}
\end{equation}
with constant weights $w_{t,x}$ and $w_{f,x}$. For the trapped contribution we
adopt a Fermi-Dirac distribution with temperature and degeneracy parameter
denoted as $T_{t,x}$ and $\eta_{t,x}$:
\begin{equation}
  g_{t,x}(k) =
  \left[e^{\left(k/T_{t,x}-\eta_{t,x}\right)} + 1\right]^{-1} \,,
  \label{eq:g_nu_thick}
\end{equation}
while the free-streaming one is modeled as a generalized Maxwell-Boltzmann
distribution with temperature $T_{f,x}$, pinching parameter $c_f$, and
normalization factor $\beta_{f,x}$:
\begin{equation}
  g_{f,x}(k) =
  \beta_{f,x} ~ k^{c_{f,x}} \, e^{-k/T_{f,x}} \,.
  \label{eq:g_nu_thin}
\end{equation}
We fix $c_f=0.6$ for all neutrino species following the fits presented in
Ref.~\cite{Tamborra:2012ac} in the context of \acp{CCSN}.
No prescription is made regarding the functional form of the angular part of the
distribution function, we only require that it satisfies $\int \dd\Omega \,
h_x(\Omega) = 4\pi$. This relation is fulfilled exactly both in the trapped and
free-streaming regimes, where $h(\Omega)$ is either constant or forward peaked
along the direction of propagation. It is only approximately fulfilled in a
generic situation in between the two extremes. This simplification allows us to
avoid dealing with complex and computationally expensive quadratures over the
angular variables in the computation of emissivities and opacities.

In the following we detail how the parameters appearing in the distribution
function are set. Two options are considered: reconstruction from \mone
quantities and equilibrium with the fluid.

\paragraph{Reconstruction from gray \mone neutrino quantities}

The Eddington factor is
used to weight the trapped and free-streaming parts in \eqref{g_nu_total} in the
same way as in the Minerbo closure \cite{Minerbo:1978a}, \ie, $w_{f,x} =
(3\chi_x-1)/2$ and $w_{t,x} = 3(1-\chi_x)/2$. 
To reconstruct the parameters entering the trapped component, we require it to
satisfy Eqs.~(\ref{eq:m1_zeroth_number_moment}) and
(\ref{eq:m1_zeroth_energy_moment}), resulting in
\begin{align}
  n_x = \frac{4\pi}{(hc)^3} \, T_{t,x}^3 F_2(\eta_{t,x}) \, ,
  \label{eq:n_m1_thick} \\
  J_x = \frac{4\pi}{(hc)^3} \, T_{t,x}^4 F_3(\eta_{t,x}) \, ,
  \label{eq:J_m1_thick}
\end{align}
where complete Fermi-Dirac integrals of order $p$ are denoted as $F_p$.
The ratio between Eqs.~(\ref{eq:J_m1_thick}) and (\ref{eq:n_m1_thick}) gives $T_{t,x}$
as a function of $\eta_{t,x}$:
\begin{equation}
  T_{t,x} = \frac{J_x}{n_x} ~ \frac{F_2(\eta_{t,x})}{F_3(\eta_{t,x})}\,.
  \label{eq:T_m1_thick}
\end{equation}
Plugging \eqref{T_m1_thick} into \eqref{n_m1_thick}, one gets the relation
\begin{equation}
  1 = \frac{4 \pi }{(hc)^3} \frac{J_x^3}{n_x^4}
  \frac{\left[F_2\left(\eta_{t,x}\right)\right]^4}
  {\left[F_3\left(\eta_{t,x}\right)\right]^3}\,,
\end{equation}
which can be inverted to find $\eta_{t,x}$. We approximate the inverse of
$\tilde{F}(\eta) = \left[F_2\left(\eta\right)\right]^4/\left[F_3\left(\eta\right)\right]^3$
as a piecewise high-order rational function split into three different intervals,
achieving an accuracy with a relative error of at most $10^{-2}$ over the values
of interest, \ie, $\eta_t\in[-5, 5]$. Once $\eta_{t,x}$ is known,
\eqref{T_m1_thick} returns the value of $T_{t,x}$. We evaluate Fermi-Dirac integrals
using the fast and accurate approximated formulas provided by
Ref.~\cite{Fukushima:2015a}.

In the free-streaming regime, the same line of reasoning results in the zeroth
number and energy moments being written in terms of the Gamma function
$\Gamma(x)$:
\begin{align}
  n_x = \frac{4\pi\beta_{f,x}}{(h c)^3}\,T_{f,x}^{c_f+3} \, \Gamma(c_f+3) \,,
  \label{eq:n_m1_thin} \\
  J_x = \frac{4\pi\beta_{f,x}}{(h c)^3}\,T_{f,x}^{c_f+4} \, \Gamma(c_f+4) \,.
  \label{eq:J_m1_thin}
\end{align}
By applying the identity $x=\Gamma(x+1)/\Gamma(x)$ to the ratio between Eqs.~
(\ref{eq:J_m1_thin}) and (\ref{eq:n_m1_thin}), the value of the
temperature is reconstructed as
\begin{equation}
  T_{f,x} = \frac{J_x}{n_x(c_f+3)} \, .
  \label{eq:T_m1_thin}
\end{equation}
Then, by combining Eqs.~(\ref{eq:n_m1_thin}) and (\ref{eq:T_m1_thin}), we recover the value
of $\beta_{f,x}$ as well:
\begin{equation}
  \beta_{f,x}  = \frac{(hc)^3 n_x}{4 \pi~\Gamma(c_f + 3)}
  \frac{1}{T_{f,x}^{c_f + 3}} \,.
\end{equation}

Note that although the parameters have been reconstructed separately for the
trapped and free-streaming parts, the condition $w_{t,x}+w_{f,x}=1$ ensures that
also the full $g_x$ satisfies Eqs.~(\ref{eq:m1_zeroth_energy_moment}) and
(\ref{eq:m1_zeroth_number_moment}). 
Moreover, the neutrino energy flux
density $H^\alpha_x$ is not required for the reconstruction, since we do not need
to model explicitly the angular part of $f_x$.

\paragraph{Equilibrium distribution function with optically thin correction}

If we assume the distribution function to describe equilibrium with the fluid,
we set $w_{f,x}=0$ and $w_{t,x}=1~\forall x$. The distribution function is then
a Fermi-Dirac function with parameters set by thermal and chemical equilibrium
as
\begin{align}
  T_{t,x} &= T \,,
            \label{eq:T_eq_thick} \\
  \eta_{t,\nue} &= \frac{\mu_{\p} - \mu_{\n} + \mu_{\tn{e}}}{T} \,,
                   \label{eq:eta_eq_thick} \\
  \eta_{t,\anue} &= - \eta_{t,\nue} \,, \\
  \eta_{t,\nux} &= 0 \,.
                   \label{eq}
\end{align}

The equilibrium assumption breaks down once neutrinos and matter decouple from
each other. In order to account for it, one can correct the energy-integrated
absorption and scattering opacities by the following multiplicative factor
\cite{Foucart:2016rxm,Radice:2021jtw}:
\begin{equation}
  C_{\tn{thin}} = \left(\frac{\varepsilon_x}{\varepsilon_{\tn{eq},x}}\right)^2 \, ,
  \label{eq:opt_thin_corr_factor}
\end{equation}
where $\varepsilon_x\equiv J_x/n_x$ is the local average neutrino energy, as evolved
by the gray \mone transport scheme, while
$\varepsilon_{\tn{eq},x}$ is the one at thermodynamic equilibrium. This choice
is motivated by the fact that the interactions taken into account have cross
sections scaling as $T^2$. Equation~(\ref{eq:opt_thin_corr_factor}) is close to $\sim 1$ as
long as the equilibrium assumption is justified, while in optically thin
conditions it is able to approximately reproduce the opacities as obtained from
the local neutrino quantities, as shown in \secref{E2_vs_nu_f}.

\section{Postprocessing strategy}
\label{sec:postproc_strategy}

\subsection{Simulation overview}
\label{sec:simulation_overview}

We consider an equal mass \ac{BNS} system with \ac{NS} masses of $\unit[{\sim}
1.298]{M_{\odot}}$. We numerically evolve it through the inspiral, merger, and
postmerger phases using the gray moment-based general-relativistic
neutrino-radiation hydrodynamics code \textsc{THC\_M1} \cite{Radice:2012cu,
Radice:2013xpa, Radice:2013hxh, Radice:2021jtw}, which is based on the Einstein
Toolkit \cite{Loffler:2011ay, Etienne:2021zndo}. The relevant neutrino
reactions used in this code and their implementation are detailed in
Ref.~\cite{Radice:2018pdn}. While there is a good overlap with the reactions
described in \secref{neutrino_reactions} (with the notable exceptions of the
scattering off electrons/positrons and nucleon decays), we stress that they differ
substantially from the ones in \bnsnurates in terms of accuracy and actual
implementation. This is, however, not relevant for this work, since the
simulation outcome will be only used to provide suitable fluid and radiation
conditions to test the new library. Spacetime is consistently evolved by the
\textsc{CTGamma} code \cite{Pollney:2009yz, Reisswig:2013sqa}, which solves
the Z4c formulation of Einstein’s field equations \cite{Bernuzzi:2009ex,
Hilditch:2012fp}.

We use the \textsc{Carpet} adaptive mesh refinement driver \cite{Schnetter:2003rb,
Reisswig:2012nc}, implementing the Berger–Oilger scheme with refluxing
\cite{Berger:1984zza, Berger:1989a}. In particular, the evolution grid employs
seven levels of refinement and the spatial resolution is halved at each finer
grid. The merging binary is simulated at a resolution for which the spacing of
the finest grid is equal to $\Delta x\unit[\approx 247]{m}$.
Within such a scheme, during the inspiral, mass conservation is achieved  with a typical relative
accuracy of ${\sim}10^{-11}$ per
iteration, on average positive. After merger, the relative variation
per iteration increases (${\sim}10^{-9}$), possibly also due to failures in the primitive variable
reconstructions both in the dense and dilute portions of the computational domain.
The total increase in mass is partially compensated by the ejecta leaving the 
computational domain once the ejection of matter from the system sets off.

To close the system of hydrodynamics equations and to provide the density-,
temperature-, and composition-dependent quantities required by the gray neutrino
transport, we adopt the Hempel and Schaffner-Bielich (Density Dependent 2)
\ac{EOS} \cite{Typel:2009sy, Hempel:2009mc} (hereafter DD2 \ac{EOS}). This is a
rather stiff \ac{EOS} that predicts a maximum nonrotating \ac{NS} mass of $\unit[2.42]
{M_{\odot}}$ and a radius $R_{1.4} = \unit[13.2]{km}$ for a nonrotating $\unit[1.4]{M_{\odot}}$
\ac{NS}. We employ this \ac{EOS} since it generally provides
long-lived merger remnants, especially in the case of light \ac{BNS} systems.

To account for turbulent viscosity of magnetic origin, we employ the
general-relativistic large-eddy simulation
formalism \cite{Radice:2017zta, Radice:2020ids}. Within this formalism,
turbulent viscosity is parametrized in terms of a density-dependent prescription
for the characteristic length scale, which was calibrated using data from the
high-resolution general-relativistic magnetohydrodynamics simulations of
Refs.~\cite{Kiuchi:2017zzg, Kiuchi:2022nin,Radice:2023zlw}.

Irrotational, $T\simeq0$, neutrinoless weak-equilibrium initial data are
constructed using the pseudospectral code \textsc{Lorene}
\cite{Gourgoulhon:2000nn}, selecting an initial separation of $\unit[45]{km}$.

\subsection{Calculation of emissivity and opacity}

\begin{table*}
  \centering
\begin{tabularx}{\textwidth}{>{\centering\arraybackslash}X >{\centering\arraybackslash}X >{\centering\arraybackslash}X >{\centering\arraybackslash}X m{12em} >{\centering\arraybackslash}X >{\centering\arraybackslash}X >{\centering\arraybackslash}X >{\centering\arraybackslash}X}
\toprule
\makecell{Name \\ ~} & \makecell{$\rho$ \\ $[\unit[]{g\,cm^{-3}}]$} & \makecell{$T$ \\ $[\unit[]{MeV}]$} & \makecell{$Y_{\rm e}$ \\ ~} & \makecell[l]{($Y_{\nue}$, $Y_{\anue}$, $Y_{\nux}$) \\ ($Z_{\nue}$, $Z_{\anue}$, $Z_{\nux}$) $[\unit[]{MeV}]$ \\ ($\chi_{\nue}$, $\chi_{\anue}$, $\chi_{\nux}$)} & \makecell{$\Delta U$ \\ $[\unit[]{MeV}]$} & \makecell{($m_{\n}^*$, $m_{\p}^*$) \\ $[10^{2} ~ \unit[]{MeV}]$} & \makecell{$\Hat{\mu}$ \\ $[\unit[]{MeV}]$} & \makecell{$\mu_{\tn{e}}$ \\ $[\unit[]{MeV}]$} \\
\midrule 
A & $6.90 \times 10^{14}$ & 12.39 & 0.07 & \makecell[l]{ $(0.09, 2.93, 0.54) \times 10^{-4}$ \\ $(0.03, 1.29, 0.21) \times 10^{-2}$ \\ $(0.33,0.33,0.33)$} & $1.90 \times 10^{1}$ & (2.81,2.80) & $2.10 \times 10^{2}$ & $1.87 \times 10^{2}$ \\[16pt]
B & $9.81 \times 10^{13}$ & 16.63 & 0.06 & \makecell[l]{ $(0.36, 2.21, 0.92) \times 10^{-3}$ \\ $(0.18, 1.23, 0.48) \times 10^{-1}$ \\ $(0.33,0.33,0.33)$} & $3.48 \times 10^{1}$ & (7.36,7.35) & $9.90 \times 10^{1}$ & $8.23 \times 10^{1}$ \\[16pt]
C & $9.87 \times 10^{12}$ & 8.74 & 0.06 & \makecell[l]{ $(0.82, 2.01, 0.84) \times 10^{-3}$ \\ $(0.22, 0.57, 0.23) \times 10^{-1}$ \\ $(0.33,0.33,0.33)$} & 4.98 & (9.16,9.15) & $4.11 \times 10^{1}$ & $3.65 \times 10^{1}$ \\[16pt]
D & $1.00 \times 10^{12}$ & 6.61 & 0.11 & \makecell[l]{ $(0.72, 0.45, 0.15) \times 10^{-2}$ \\ $(1.51, 0.93, 0.39) \times 10^{-1}$ \\ $(0.33,0.33,0.33)$} & 0.47 & (9.37,9.36) & $1.76 \times 10^{1}$ & $1.94 \times 10^{1}$ \\[16pt]
E & $1.00 \times 10^{11}$ & 3.60 & 0.14 & \makecell[l]{ $(10.92, 7.88, 0.29) \times 10^{-3}$ \\ $(12.48, 10.21, 0.83) \times 10^{-2}$ \\ $(0.34,0.37,0.36)$} & $0.45 \times 10^{-1}$ & (9.39,9.38) & 8.44 & 9.05 \\[16pt]
F & $1.01 \times 10^{10}$ & 2.17 & 0.19 & \makecell[l]{ $(12.56, 13.00, 0.36) \times 10^{-3}$ \\ $(11.89, 15.56, 0.96) \times 10^{-2}$ \\ $(0.53,0.57,0.59)$} & $0.72 \times 10^{-3}$ & (9.40,9.38) & 4.57 & 4.17 \\
\bottomrule
 \end{tabularx}
\caption{Properties of the representative grid points A$-$F extracted from the 
simulation at
$t=\unit[50]{ms}$ after merger. $\rho$ represents the rest mass density, $T$ the
fluid temperature, $\ye$ the electron fraction, $Y_x$, $Z_x$, and $\chi_x$ (with
$x \in \{\nue, \anue, \nux \}$) the neutrino number fractions, energy fractions
and Eddington factors, respectively, $\Delta U$ the neutron-to-proton \ac{RMF}
interaction potential difference, $(m_{\n}^*, m_{\p}^*)$ the effective nucleon
masses, $\hat{\mu}$ the neutron-to-proton chemical potential difference, and
$\mu_{\tn{e}}$ the electron chemical potential. Please note that here the chemical potentials are
the relativistic ones, \ie, they include the rest mass contribution.}
 \label{tab:thermodynamics_points}    
\end{table*}

\begin{figure}
  \centering
  \includegraphics[width=\columnwidth]{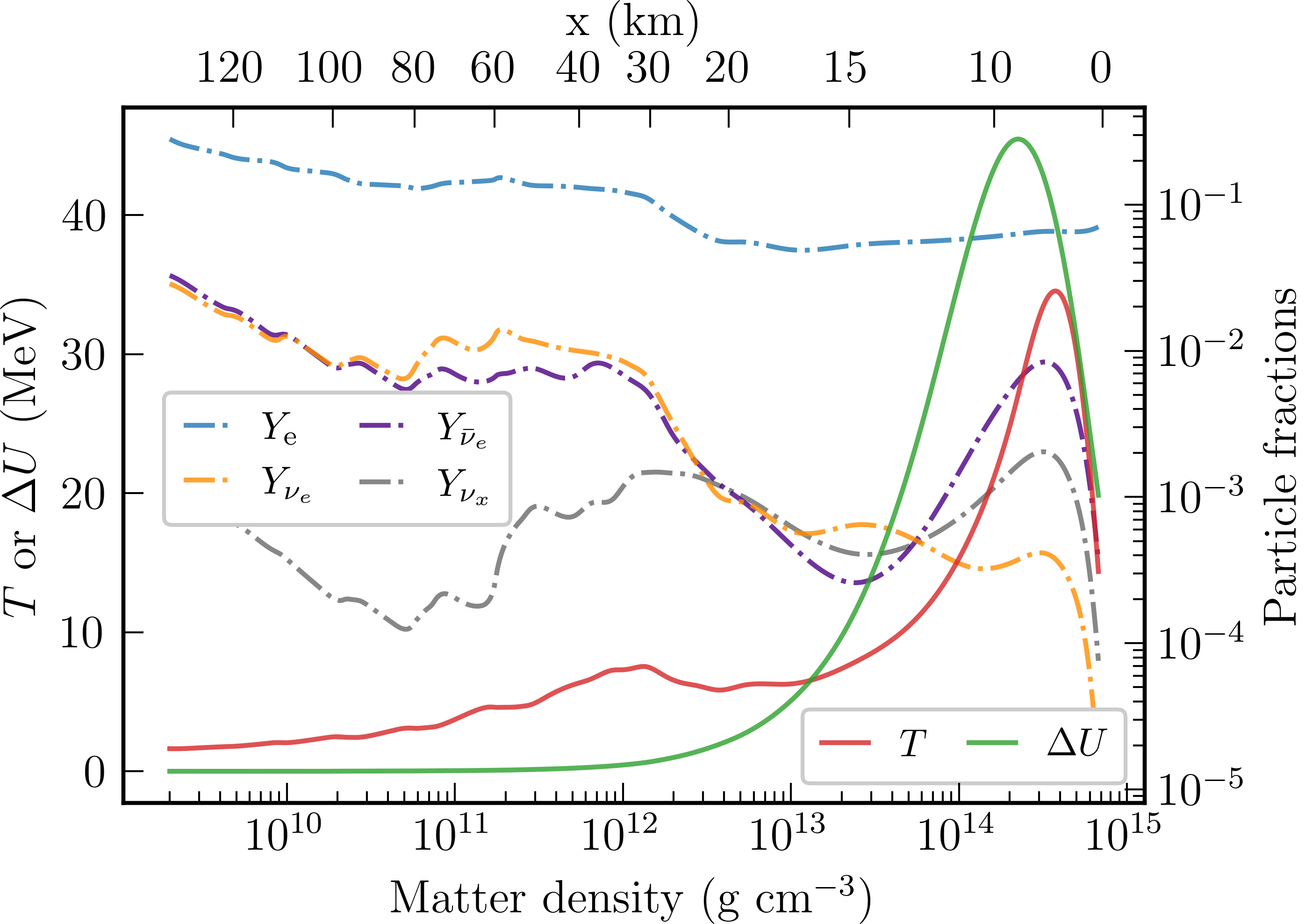}
  \caption{Profiles of the matter temperature, $\Delta U$, electron fraction, and
    neutrino abundances as a function of the rest mass density as extracted from
    the simulation along the positive $x$ axis.}
  \label{fig:thermo_DD2_M12980-12980_M1_LR_axis_x_t_12380.8_rl_3}
\end{figure}

\begin{figure}
  \centering
  \includegraphics[width=\columnwidth]{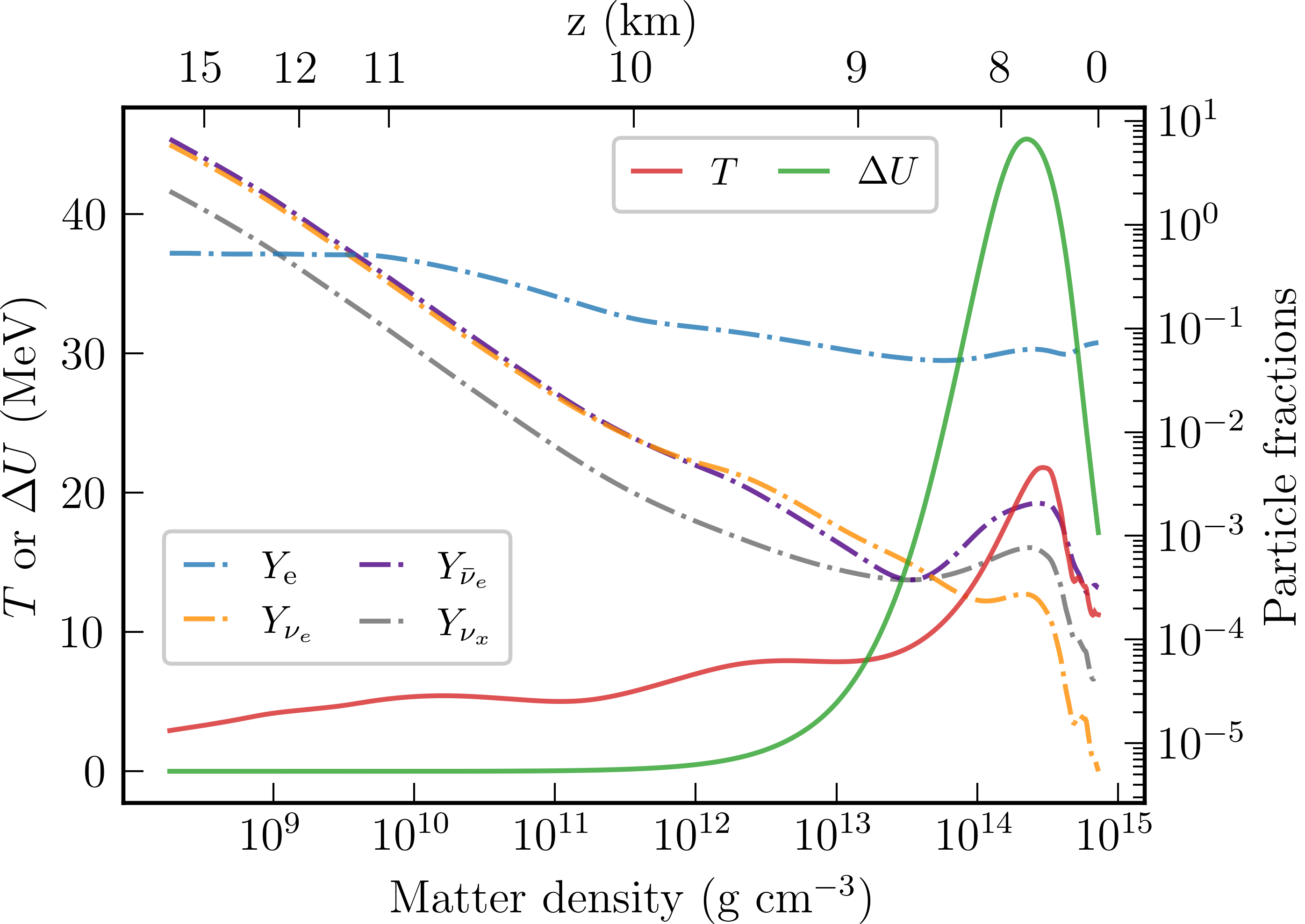}
  \caption{Same as
    \figref{thermo_DD2_M12980-12980_M1_LR_axis_x_t_12380.8_rl_3}, but along the
    positive $z$ axis.}
  \label{fig:thermo_DD2_M12980-12980_M1_LR_axis_z_t_12380.8_rl_6}
\end{figure}

To compute neutrino emissivities and opacities in conditions that are relevant
for \ac{BNS} mergers, we extract them from the simulation
presented in \secref{simulation_overview}, at $t=\unit[50]{ms}$ after merger. At
this point in the evolution,
the remnant is approximately axisymmetric and it has reached a quasistationary
state, further evolving on the longer cooling and viscous timescales. In our
postprocessing procedure, we consider (i) single grid points, (ii) one-dimensional
slices, as well as (iii) the whole three-dimensional computational domain.

In the first case, we define a set of reference rest mass densities, $\left\{
\rho_i \right\}$, including a value close to the maximum rest mass density
inside the remnant (point A) and several other densities almost equally spaced in
logarithmic space (points B$-$F). For each of them, we compute the volume-averaged
density, $\langle \rho_i \rangle$, temperature, $\langle T_i \rangle$, and
electron fraction, $\langle Y_{e,i} \rangle$, evaluated over all the cells with density $\rho$
satisfying the condition $ | \rho - \rho_i |/ \rho_i \leq 0.05$ and within an angular
slice of width $\Delta \theta {\sim} 20^{\circ}$ across the
equatorial plane. After that, we search for the point
inside the same region whose conditions are the closest to the averaged ones by
minimizing the Euclidean distance with respect to $(\log\langle \rho_i
\rangle,\langle T_i \rangle,\langle Y_{e,i} \rangle)$. This procedure ensures
that the selected points are representative. \tabref{thermodynamics_points}
summarizes the conditions of the selected points and their most relevant
properties.

For 1D slices, starting from the center of the computational domain, we consider
all the cells along the positive $x$ or $z$ axis, for a fixed refinement level.
In Figs.~\ref{fig:thermo_DD2_M12980-12980_M1_LR_axis_x_t_12380.8_rl_3} and
\ref{fig:thermo_DD2_M12980-12980_M1_LR_axis_z_t_12380.8_rl_6} we present the
profiles of the temperature, the \ac{RMF} potential energy shift, $\Delta U$
\footnote{The \ac{EOS} employed predicts $\Delta m^*=\Delta m$ for any condition, therefore the information
about \ac{RMF} effects is
completely encoded into the value of $\Delta U$.},
and the
relevant particle abundances as a function of the rest mass density along the
positive $x$ and $z$ axis, respectively.

We instead consider the entire three-dimensional domain when we compute
the position of the neutrino surfaces (see \secref{neutrino_surface}). In
this case, the computational grid extends over the entire region covered 
by the coarsest refinement level and is defined in terms of spherical coordinates with
a log-spaced radial mesh and equally spaced angular bins.

In all three cases, the data that we extract from the simulation are the
thermodynamic conditions $(\rho,T,\ye)$ (necessary to call the \ac{EOS}), as
well as the neutrino quantities evolved by the \mone scheme, \ie, $n_x$, $J_x$,
and the Eddington factor, $\chi_x$, for all the relevant neutrino
species. 
We then employ the neutrino quantities to reconstruct the neutrino distribution
functions, as detailed in \secref{nu_distribution}.
Once the conditions are set, we use the \bnsnurates library to compute both
energy-dependent and energy-integrated emissivities and opacities for all
neutrino species and for all the reactions presented in
\secref{neutrino_reactions}.
Note that the simulation just considers three neutrino species,
\ie, $\left\{\nue,\anue,\nux\right\}$, as the employed neutrino scheme does not
involve any reaction that distinguishes between heavy-type neutrinos
and antineutrinos. On the other hand, in \bnsnurates, the implementation
of the inelastic neutrino scattering off $e^\pm$ introduces that sort of
distinction. Nonetheless, as the difference is only marginal,
in Secs.~\ref{sec:spectral} and \ref{sec:gray} we just present the results for $\nux$ neutrinos.

\subsection{Neutrino surface calculations}
\label{sec:neutrino_surface}

The neutrino optical depth, $\tau_x$, quantifies the number of interactions
experienced by diffusing neutrinos along a certain path. Denoting as
$P_B(\boldsymbol{x}; t)$ the set of all paths that reach the boundary of our
computational domain at time $t$ starting from the point $\boldsymbol{x}$, 
the optical depth can be defined as
\begin{equation}
  \label{eq:opt depth definition}
  \tau_{x}(\boldsymbol{x}; t) := \min_{\Gamma \in P_B(\boldsymbol{x}; t)} \int_\Gamma
  \lambda^{-1}_{x}\,\dd s \, .
\end{equation}
We then define neutrino surfaces as the isosurfaces where $\tau_{x}$ attains the
value 2/3. Energy-dependent inverse mean free paths result in energy-dependent
optical depths and neutrino surfaces. Depending on the definition of
$\lambda^{-1}_x$ entering in \eqref{opt depth definition}, two types of optical
depths are usually considered: the diffusion optical depth
($\tau_{\tn{diff},x}$) and the equilibration optical depth ($\tau_{\tn{eq},x}$).
In the former case, one has that
\begin{equation}
  \label{eq:diff_inv_mfp}
  \lambda^{-1}_{\tn{diff},x} \equiv \sum_l \lambda^{-1}_{l,x} \, ,
\end{equation}
where the sum runs over all processes providing opacity, and the neutrino
surfaces correspond to the last interaction surface for a neutrino moving from
optically thick to optically thin regions. In the second case, one has instead
that
\begin{equation}
  \label{eq:eq_inv_mfp}
  \lambda^{-1}_{\tn{eq},x} \equiv \sqrt{ \left( \sum_{l'}
      \lambda^{-1}_{{l'},x} \right) \lambda^{-1}_{\tn{diff},x} }
\end{equation}
and the sum runs on inelastic processes only. In this case, the neutrino
surfaces enclose the portion of the system in which neutrinos are in thermal
contact with matter.

We consider a snapshot of the 3D computational domain as extracted from our
simulation at $t=\unit[50]{ms}$ after merger and we compute
energy-dependent optical depths and neutrino surfaces using \eqref{opt depth definition},
relying on the algorithm depicted in \appref{tau_nusphere}. The neutrino surfaces
presented in Secs.~\ref{sec:NES_NPS} and \ref{sec:energy_shift} are isosurfaces of the
equilibration optical depth, $\tau_{\tn{eq},x}$.

\section{Results: energy dependent rates}
\label{sec:spectral}

\subsection{Reaction comparison}
\label{sec:reaction_comparison}

\begin{figure*}
  \centering
  \includegraphics[width=\textwidth]{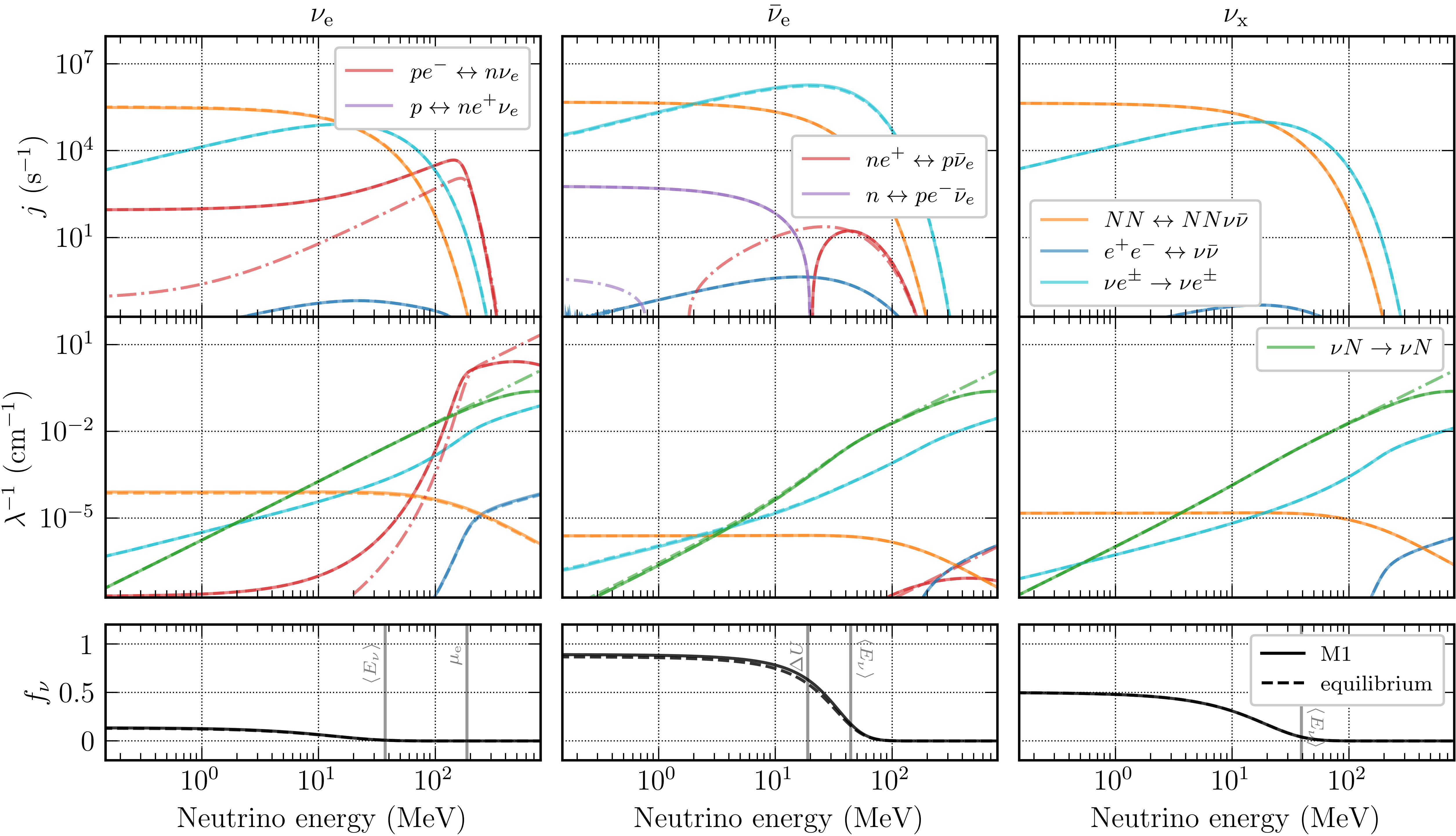}
  \caption{Top (middle) row: spectral emissivity (inverse mean free path) for
    different reactions and for the thermodynamic conditions of point A in
    \tabref{thermodynamics_points}. Bottom row: neutrino occupation number as a
    function of the neutrino energy, superimposed to some relevant energy scales. Left, center, and right panels refer to
    electron neutrinos, electron antineutrinos, and heavy-type neutrinos,
    respectively.
    Solid lines were obtained by using the reconstructed distribution functions, while dashed lines the equilibrium ones. 
    In both cases, $\beta$ processes account for weak magnetism and \ac{RMF} effects 
    and isoenergetic scattering reactions for weak magnetism.
    Red, purple, and green dot-dashed lines differ from solid ones for the absence of such effects.}
  \label{fig:DD2_M12980-12980_M1_LR_t_12390.4_n_50_rho_7.0e+14}
\end{figure*}

\begin{figure*}
  \centering
  \includegraphics[width=\textwidth]{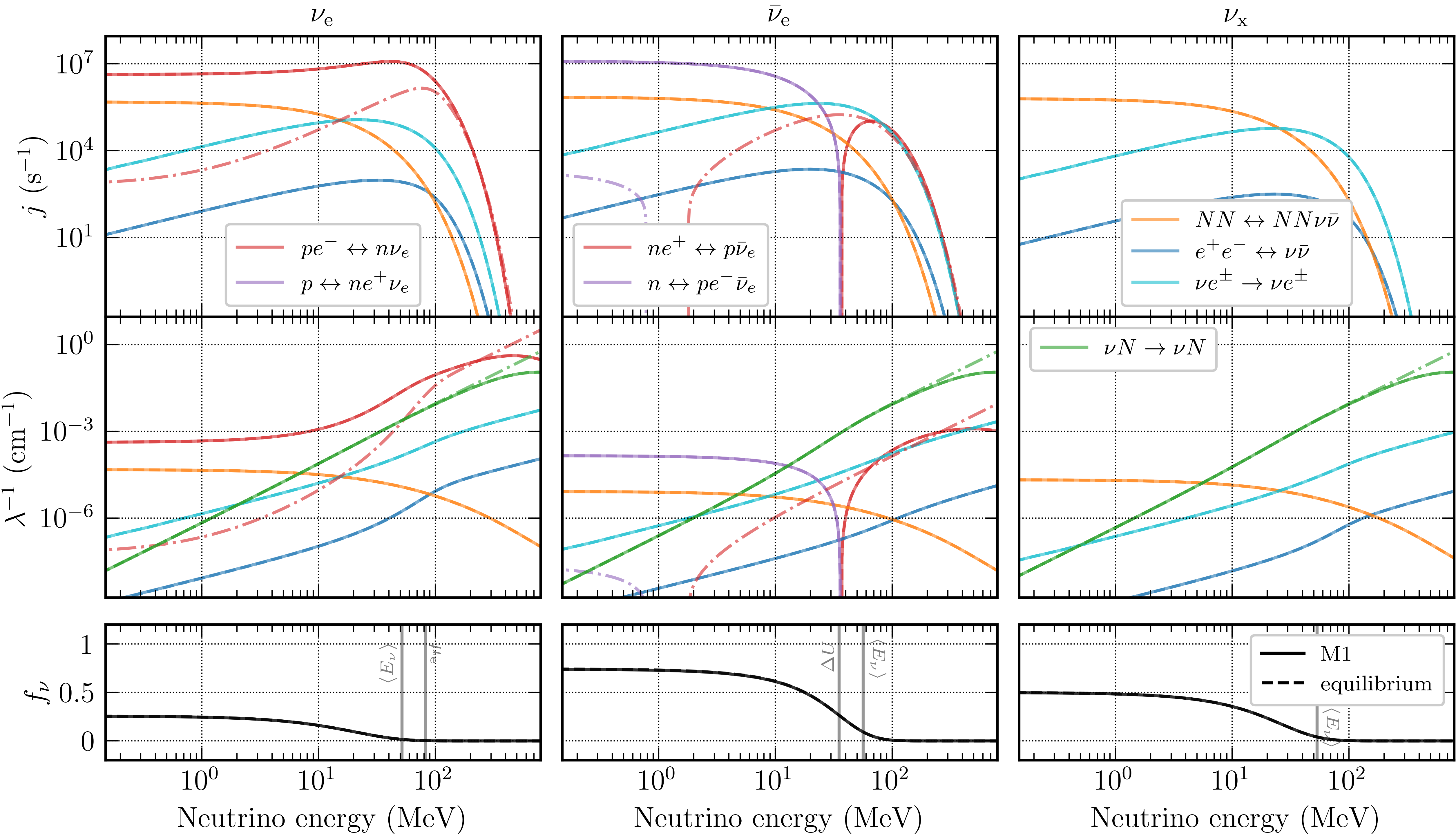}
  \caption{Same as
    \figref{DD2_M12980-12980_M1_LR_t_12390.4_n_50_rho_7.0e+14}
    but for thermodynamic point B in \tabref{thermodynamics_points}.}
  \label{fig:DD2_M12980-12980_M1_LR_t_12390.4_n_50_rho_1.0e+14}
\end{figure*}

In Figs.~\ref{fig:DD2_M12980-12980_M1_LR_t_12390.4_n_50_rho_7.0e+14} and
\ref{fig:DD2_M12980-12980_M1_LR_t_12390.4_n_50_rho_1.0e+14}
we show spectral emissivities (top panels) and  inverse mean free paths (middle
panels) for different reactions, as well as the 
neutrino distribution functions (bottom panels), under the thermodynamic conditions
characterizing points A and B in \tabref{thermodynamics_points}, respectively, and over the neutrino energy
interval relevant for \ac{BNS} merger conditions. 
We recall that point A was chosen as a representative point for the conditions
in the densest part of the remnant. 
Solid lines were obtained by using the reconstructed distribution functions, while dashed lines the equilibrium ones.
The scattering off nucleons appears only in the middle panels since, due to its elastic nature, it
does not affect the distribution of the neutrino energy but it still
significantly contributes to the opacity of neutrinos.
Indeed, it is the dominant contribution to the
inverse mean free path in many regimes, especially at $k \gtrsim \unit[10]{MeV}$.

Inelastic neutrino interactions at points A and B are distinguished by a qualitative difference,
associated with the relative importance of $\beta$ processes for
electron-type (anti)neutrinos. In the center of the remnant, the contribution of
both (inverse) electron and positron captures and (inverse) nucleon decays, shown in red and purple
lines respectively, are suppressed either by
the high degeneracy of neutrons and electrons or by the paucity of positrons,
depending on the direction in which the reactions proceed. Only absorptions on neutrons of
highly energetic $\nue$ ($k_{\nue} \gtrsim \unit[100]{MeV}$) are
relevant, as they are able to produce electrons in the final state above the
Fermi level ($\mu_{\tn{e}} \approx \unit[190]{MeV}$). As a consequence, the
neutrino dynamics in the center of the remnant is established by other
reactions, similarly to what happens generally in the case of heavy-type
neutrinos. For example, the (inverse) NN bremsstrahlung, associated with
orange lines, is the most relevant
process in terms of emission (absorption) of soft neutrinos of any flavor.
Inelastic \ac{NEPS} is also an important process for neutrinos in the center of
the remnant (cyan lines). High-energy neutrinos have a larger probability to scatter off
energetic electrons and excite them above the degeneracy level. Assuming that
neutrino distribution functions are close to the equilibrium ones,
as shown in the bottom panels,
\ac{NEPS} efficiently produces neutrinos in the final state with a broad energy
spectrum. While for very soft neutrinos its contribution is always subdominant
with respect to NN bremsstrahlung, it becomes very relevant for $k
\gtrsim \unit[20]{MeV}$ or even lower energies for $\anue$.

Moving out from the center of the remnant, the density decreases and
charged-current reactions are no longer inhibited. According to the left panel of
\figref{DD2_M12980-12980_M1_LR_t_12390.4_n_50_rho_1.0e+14}, in the region where point B is located, electron captures and
absorption on neutrons are the dominant reactions undergone by $\nue$ neutrinos
of any energy, once microphysics effects are accounted for
(\cf solid red lines).
Electron captures on protons retain an almost
constant emissivity ($j {\sim} \unit[10^7]{s^{-1}}$) until several tens of MeV,
before dropping off as a consequence of the low occupation of
highly energetic electrons above the Fermi level. At the same time, the mean free path associated with
$\nue$ absorptions on neutrons is well below the size of the remnant (a few ${\sim} \unit[10^{6}]{cm}$)
for any
neutrino energy. Other contributing processes for the electron neutrino dynamics
are the scattering off nucleons and the (inverse) NN bremsstrahlung,
with the latter that partially increases the overall emissivity and inverse
mean free path for soft enough $\nue$ (${\sim} 10\%$ increase for
$k \lesssim \tn{a few MeV}$). Other reactions are instead subdominant in
these conditions, such as (inverse) $\ep \, \em$ annihilations (blue lines),
inhibited for any flavor by the electron degeneracy at high densities, and (inverse)
proton decays, which turn out to be kinematically forbidden for any conditions
considered in this study, particularly when considering the impact on its kinematics due to \ac{RMF} effects. Speaking
of such effects, the central panels of
\figref{DD2_M12980-12980_M1_LR_t_12390.4_n_50_rho_1.0e+14}
highlight their importance when considering electron antineutrino interactions.
Because of kinematics constraints, charged-current interactions are
possible only below or above a given neutrino energy threshold. Under the
assumption of zero-momentum transfer, the threshold energies for (inverse)
positron captures and (inverse) neutron decays are separated by an energy gap of $2
m_{\tn{e}} \unit[{\sim} 1]{MeV}$, whose position in the spectrum is ultimately
determined by the average nucleon energy difference. Therefore, when \ac{RMF} effects
are taken into account, the values of the energy thresholds are modified from $\Delta m \pm m_{\rm e}$
to $\Delta m^* + \Delta U \pm m_{\rm e}$.
Since \ac{RMF} effects are sizable for high enough densities (see Figs.~
\ref{fig:thermo_DD2_M12980-12980_M1_LR_axis_x_t_12380.8_rl_3} and
\ref{fig:thermo_DD2_M12980-12980_M1_LR_axis_z_t_12380.8_rl_6}), as in the case of
point B, the kinematic thresholds are effectively shifted to higher energies
($\Delta U {\sim} \unit[35]{MeV}$). This confines $\ep$
captures and absorptions on protons to be effective only for $\anue$ with $k
\gtrsim$ several tens of MeV. At the same time, the influence of (inverse)
neutron decays extends over a broader energy range and becomes significant for
determining the behavior of soft electron antineutrinos. It is also interesting
to notice how the contribution from $\anue\,\tn{e}^\pm$ scattering helps 
to make the energy dependence of the total $j_{\anue}$ and $\lambda_{\anue}^{-1}$ smoother,
by partially filling the void between the two energy
thresholds due to electron antineutrinos not undergoing $\beta$ processes
(which is also an artifact of our
approximated \ac{RMF} treatment with respect to more sophisticated approaches,
see \cite{Oertel:2020pcg}). In comparison to the electron neutrino case, once we combine
together the contributions of the various processes, we find that $j_{\anue}$
is dominated by neutron decays for the production of soft particles, while it is
suppressed by the lower occupation of positrons compared to electrons at higher energies
($k {\sim}$ a few tens of MeV), where $j_{\nue}$ instead peaks.
Similarly, the total inelastic $\lambda_{\anue}^{-1}$
is typically smaller than the one of $\nue$, in particular for highly energetic neutrinos.
The latter feature is a direct consequence of the partial
Pauli blocking from neutrons, which is still relevant even if less important than
for point A, and of the fact that at large enough energies $\nue$
absorptions are no longer inhibited by the degeneracy of final-state electrons.

Because of the absence of $\mu^{\pm}$ and $\tau^{\pm}$ leptons in our description of the system, $\nux$
production and removal rely on pair processes and inelastic scattering off $e^\pm$.
Therefore, heavy-type neutrinos are typically created and destroyed at comparable or smaller
rates than electron-flavored ones, especially where $\beta$ processes are
relevant. On the other hand, the contribution to the opacity of elastic
scattering reactions off nucleons is similar with respect to the one seen for the other
neutrino flavors. Focusing on inelastic processes, NN bremsstrahlung can
efficiently convert nucleon energy into $\nux\,\anux$ pairs populating the
softest part of the spectrum, and vice versa.
For individual energies ${\lesssim}\unit[20]{MeV}$, in typical conditions inside the remnant, NN bremsstrahlung guarantees an emissivity of at least ${\sim} \unit[10^5]{s^{-1}}$ and a mean free path that is smaller than the typical remnant size ($\lambda_{\nux}^{-1} \gtrsim
\unit[10^{-5}]{cm^{-1}}$). 
The contribution of $\nux\,\tn{e}^\pm$ scattering is
qualitatively similar to the case of electron (anti)neutrinos, but the impact on
the total $j_{\nux}$ and $\lambda_{\nux}^{-1}$ is more significant as there are no charged-current
reactions. In fact, it extends the upper tail of $j_{\nux}$ to slightly higher energies and,
most importantly, it significantly increases the equilibration inverse mean free path,
see \eqref{eq_inv_mfp}, ultimately affecting the decoupling conditions of $\nux$ neutrinos
from matter, as discussed in more detail in \secref{NES_NPS}.
This reaction hierarchy for heavy-type neutrinos is qualitatively similar to the one
found by Ref.~\cite{Thompson:2000gv} for similar conditions in
\acp{CCSN}.

\begin{figure*}
  \centering
  \includegraphics[width=\textwidth]{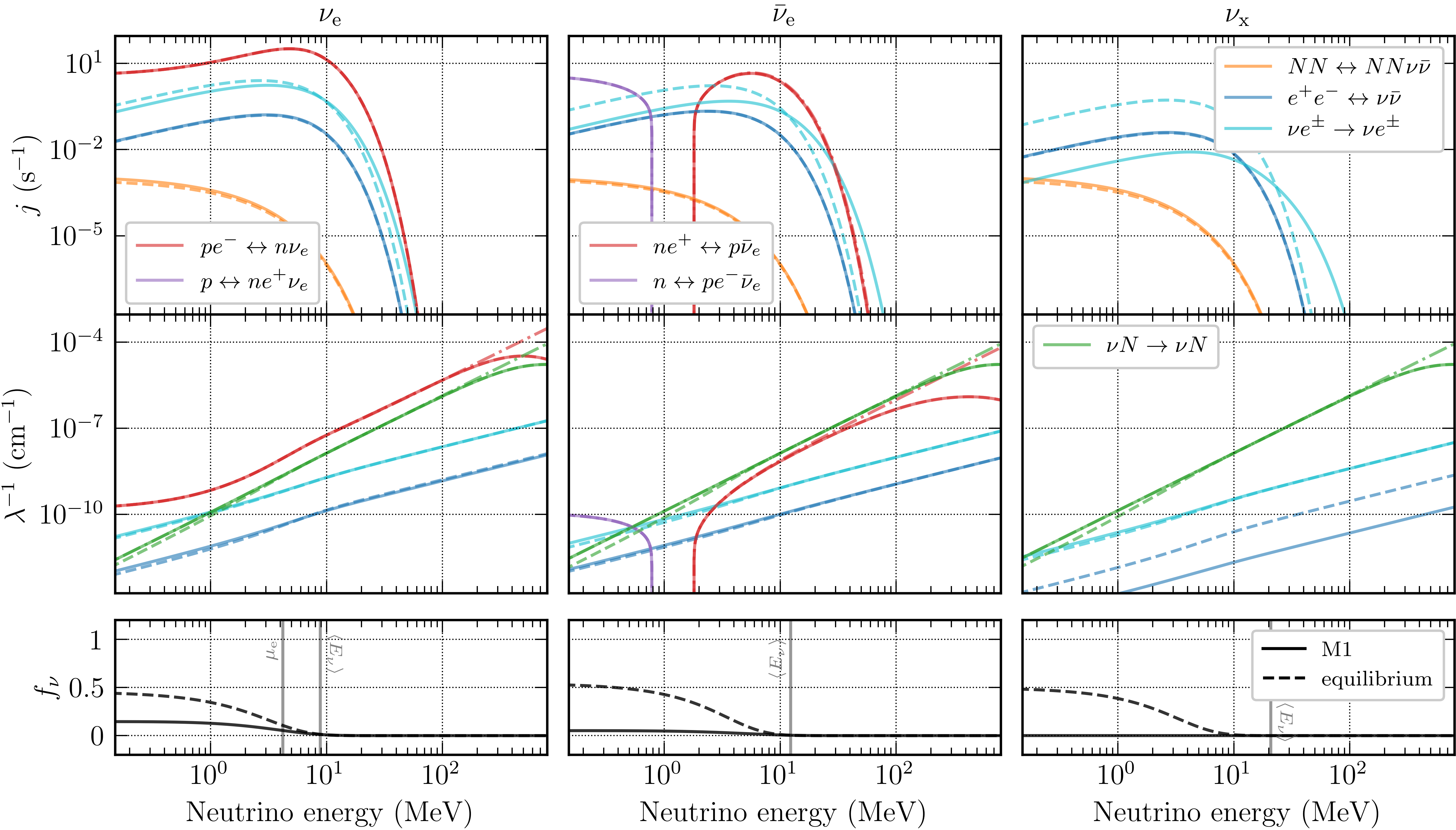}
  \caption{Same as
    \figref{DD2_M12980-12980_M1_LR_t_12390.4_n_50_rho_7.0e+14}
    but for thermodynamic point F in \tabref{thermodynamics_points}.}
  \label{fig:DD2_M12980-12980_M1_LR_t_12390.4_n_50_rho_1.0e+10}
\end{figure*}

Emissivities and inverse mean free paths tend to decrease for all
reactions once density and temperature become smaller. Since the transition
between the high and the low density regime is gradual, we avoid commenting
points at intermediate densities (C$-$E), but we include the relative figures in
\appref{additional_plots}. A detailed reaction comparison is repeated in
\figref{DD2_M12980-12980_M1_LR_t_12390.4_n_50_rho_1.0e+10}
only for the conditions extracted at point F in \tabref{thermodynamics_points}, in
order to understand how the different processes behave away from the densest parts
of the remnant, where neutrinos stream freely. Based on the simulation outcome, the neutrino number densities in this 
region are ${\sim} 2-4$ orders of magnitude smaller compared to the ones at point B.
Also the average neutrino energies are smaller ($\unit[{\sim} 10]{MeV}$ for $\nue$ and $\anue$
and $\unit[{\sim} 20]{MeV}$ for $\nux$, compared to
$\unit[{\sim} 50]{MeV}$ for all the species at point B).
Electron captures remain the
most effective channel for emitting new $\nue$ neutrinos, but the emissivity is ${\sim} 6$
orders of magnitude smaller than the one at point B for $k \lesssim \unit[10]{MeV}$.
The difference is even more pronounced above that, as the emitted
spectrum is now cut off at lower energies, due to electrons being on average less energetic
($\mu_{\tn{e}} \approx \unit[4]{MeV}$).
On the opacity side, the upper bound on the total $\lambda_{\nue}^{-1}$ is still determined mainly
by $\nue \, n$ absorptions, with a secondary contribution coming from $\nue\, N$
scattering for $k \gtrsim \unit[1]{MeV}$. Differently from what was discussed for
points at higher densities,
in these conditions the NN bremsstrahlung is suppressed for any neutrino flavor, even
for small energies, as it becomes particularly inefficient at low densities due
to its pair nature.
In the case of electron antineutrinos, we observe how 
the energy thresholds of $\beta$ processes have rolled back towards energies of
around $\unit[{\sim} 1]{MeV}$, \ie, as if \ac{RMF} effects were absent. This is to
be expected, since these effects become less significant with decreasing density
(\cf $\Delta U$ profiles in Figs.~\ref{fig:thermo_DD2_M12980-12980_M1_LR_axis_x_t_12380.8_rl_3} and
\ref{fig:thermo_DD2_M12980-12980_M1_LR_axis_z_t_12380.8_rl_6}). In the specific
conditions of point F, $\Delta U$ is negligible with respect to the bare nucleon
mass difference (see \tabref{thermodynamics_points}), therefore (inverse) neutron decays only
create (absorb) electron antineutrinos with sub-MeV energies. For the same reason, $\ep$ captures and
absorptions on protons are now relevant over a wide range of the $\anue$
spectrum, down to $\unit[{\sim} 2]{MeV}$, providing a higher emissivity 
compared to the other processes and
a mean free path that is on average closer to the one associated with $\anue \, N$
scattering reactions with respect to what was observed in Figs.~
\ref{fig:DD2_M12980-12980_M1_LR_t_12390.4_n_50_rho_7.0e+14}
and
\ref{fig:DD2_M12980-12980_M1_LR_t_12390.4_n_50_rho_1.0e+14}.

For conditions at points A and B, different assumptions on the shape of the
neutrino distribution functions do not impact on the
emissivities and inverse mean free paths, due to the fact that the reconstructed $f_{\nu}$
are very close to the equilibrium ones (see bottom panels of
Figs.~\ref{fig:DD2_M12980-12980_M1_LR_t_12390.4_n_50_rho_7.0e+14}
and
\ref{fig:DD2_M12980-12980_M1_LR_t_12390.4_n_50_rho_1.0e+14}).
Conversely, for point F, we observe qualitative differences depending on the choice of $f_\nu$.
In fact, neutrinos have already decoupled from matter in the region under
consideration, therefore assuming a distribution function at equilibrium is not
a good description of the actual neutrino occupation number in the system. Both
the emissivity for inelastic \ac{NEPS} reactions and the inverse mean free path for processes
involving $\nu\bar{\nu}$ pairs are overestimated in the case of equilibrium
$f_\nu$, as a result of the higher occupation numbers predicted for
(anti)neutrinos in the initial state. The mismatch is more evident in the case of
heavy-type neutrinos, for which the decoupling occurs at higher densities with
respect to $\nue$ and $\anue$ (see \secref{E2_vs_nu_f}). In fact, we notice that the most important
process for $\nux$ emission depends on the specific assumption made on $f_\nu$.
When considering equilibrium conditions, \ac{NEPS}
dominates the overall $\nux$ production, as the corresponding emissivity is overestimated
up to ${\sim} 2$ orders of magnitude for $k \lesssim \unit[20]{MeV}$. On the other hand,
when reconstructing $f_\nu$ from local \mone radiation quantities, $\nux$
up to $k \unit[{\sim} 10]{MeV}$ are most efficiently produced by other reactions
(\ie, NN bremsstrahlung and $\ep \, \em$ annihilations), while $\nux\,\tn{e}^\pm$
scattering dominates only at higher energies. Similar differences are observed
for the inverse mean free paths associated with
the production of $\ep\,\em$ pairs and the inverse NN bremsstrahlung. However, the $\nux$ opacity is
dominated in any case by isoenergetic scattering reactions, since inelastic processes
become irrelevant once the decoupling from matter has occurred.

\subsection{Scattering off electrons and positrons}
\label{sec:NES_NPS}

\begin{figure*}
    \centering
    \includegraphics[width = \textwidth]{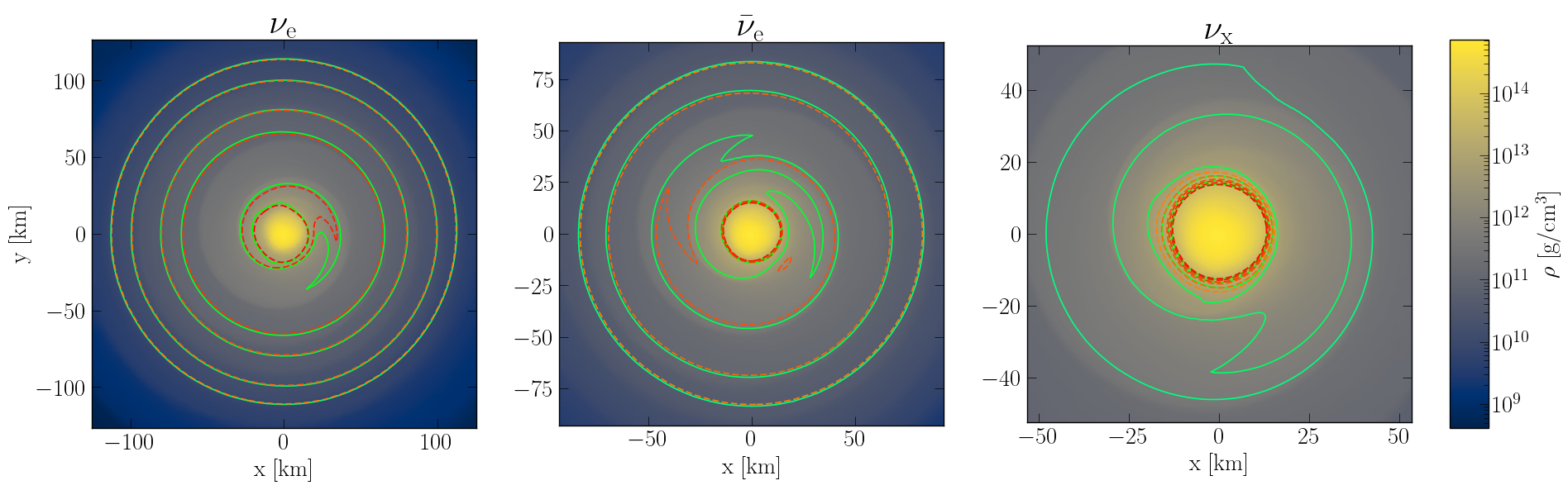}
    \caption{Energy-dependent, equilibration neutrino surfaces on the $xy$ plane,
    with (solid green) and without (dashed red) the contribution
    of inelastic scattering reactions off electrons and positrons. Color coded is the
    rest mass density on the plane.
    Each panel refers to a given neutrino species, \ie, $\nue$, $\anue$, and $\nux$, from left to right.
    Different color saturations are used to distinguish the neutrino surfaces of the following energies:
    3, 5, 7.5, 10, 17.5, and 25 $\unit[]{MeV}$, moving from inner to outer regions. Note the different spatial resolutions in the
    panels due to the different typical size of the neutrino surfaces.}
    \label{fig:DD2_M12980-12980_M1_LR-40.02-xy_plane-all_reactions-all_corrections-m1_distr-eqOpcts_True-NuSpheresCompare}
\end{figure*}

We now focus specifically on the importance of including the contribution of inelastic scattering
reactions off electrons and positrons.
In \figref{DD2_M12980-12980_M1_LR-40.02-xy_plane-all_reactions-all_corrections-m1_distr-eqOpcts_True-NuSpheresCompare}
we present the intersection between the energy-dependent neutrino surfaces and the $xy$ plane
for each of the three neutrino species.
The optical depths are evaluated using the equilibration inverse mean free path, $\lambda^{-1}_{\rm eq}$,
defined in \eqref{eq_inv_mfp}, which is obtained by either summing over all
the reactions listed in \secref{neutrino_reactions} 
(solid green)
or all the reactions except for
\ac{NEPS} (dashed red). 
We compute the neutrino surfaces for six energies: 3, 5, 7.5, 10, 17.5, and 25 $\unit[]{MeV}$, where the largest
(smallest) neutrino energy corresponds to the outermost (innermost) surface.
The surfaces are superimposed to the rest mass density in $\unit[]{g \, cm^{-3}}$ on the $xy$ plane, for reference.
In the mean free path calculations, the neutrino distribution functions are reconstructed from the gray
\mone radiation quantities. 

The left panel of \figref{DD2_M12980-12980_M1_LR-40.02-xy_plane-all_reactions-all_corrections-m1_distr-eqOpcts_True-NuSpheresCompare} shows that the effect of \ac{NEPS} on the thermalization of 
$\nue$ is marginal, as only the decoupling of neutrinos with $\unit[5]{MeV}$ is visibly affected
when including its contribution. This is a consequence of \ac{NEPS} having a strong dependence on the matter
temperature. Generally, the thermalization of $\nue$ neutrinos occurs mainly through $\beta$ processes or,
for soft neutrinos at high densities, via inverse NN bremsstrahlung.
However, when the temperature is large enough, the inverse mean free path for \ac{NEPS} can be
comparable to or larger than the one of the other reactions, as occurs in the hot annulus of matter
with $T \lesssim \unit[40]{MeV}$ (see
\figref{thermo_DD2_M12980-12980_M1_LR_axis_x_t_12380.8_rl_3}). This region
includes the outer part of the central massive remnant and the innermost part of the disk
and is where the surface for $\nue$ with $k=\unit[5]{MeV}$ is located.
In the middle panel, we notice how, for each energy and for both of the sets of reactions considered,
the electron antineutrino surfaces have smaller size compared to the electron neutrino ones.
This is mostly due to the overabundance
of neutrons with respect to protons inside the remnant,
which makes absorptions on protons less efficient that the ones on neutrons.
In fact, the position of the two outermost $\anue$ surfaces is, again, primarily determined
by $\beta$ processes, coherently
with the fact that they are pushed outward by only $\unit[{\sim} 1]{km}$ 
when considering also the $\anue e^\pm$ scattering.
On the other hand, moving deeper inside the system, we find that the inclusion of NEPS reactions significantly
expands the antineutrino surfaces with energies of $\unit[10]{MeV}$,
whose azimuthally averaged radius increases by $\unit[{\sim} 6]{km}$,
and, particularly, $\unit[7.5]{MeV}$,
which would be otherwise superimposed to the surfaces of lower energies in the absence of \ac{NEPS}.
This is explainable as, given the conditions where these
surfaces are situated (\ie, $\unit[{\sim} 10^{12}-10^{13}]{g \, cm^{-3}}$),
NEPS is the dominant inelastic process for $\anue$ opacity within the energy interval 
between a few and $\unit[{\sim} 10]{MeV}$
(see Figs.~
\ref{fig:DD2_M12980-12980_M1_LR_t_12390.4_n_50_rho_1.0e+13} and
\ref{fig:DD2_M12980-12980_M1_LR_t_12390.4_n_50_rho_1.0e+12}).
The right panel instead clearly shows that for \ac{BNS} merger conditions,
heavy-type neutrinos are the most affected
by the inclusion of \ac{NEPS} reactions. In absence of them, the possibility for $\nux$ 
to interact via inelastic reactions (typically $\beta$ processes) is greatly reduced compared to electron-flavored (anti)neutrinos,
following from the fact that the $\mu^\pm$ are not usually included in simulations and $\tau^\pm$ are not
expected to be present in the remnant.
In fact, the dashed-red neutrino surfaces of different energies are all concentrated
on the edge of the massive central remnant, where the density and temperature drop determines
the freeze-out of inverse NN bremsstrahlung and $\nux \anux$ pair annihilations.
On the other hand, when including NEPS, the volume of the region where heavy-type neutrinos
equilibrate with matter considerably increases, particularly for the two highest energies considered.
Of notable interest is the impact on the surface of $\unit[25]{MeV}$ neutrinos
(close to mean energy typically expected by current models 
for the emitted $\nux$ spectrum; see, \eg, \cite{Foucart:2015gaa,Cusinato:2021zin}), whose average radius increases
from $\unit[{\sim} 16]{km}$ to $\unit[{\sim} 45]{km}$. 
The modification of this surface is, likely, directly connected to a variation of the 
mean energy, since the decoupling from matter occurs at different thermodynamics conditions.
In particular, the $\nux$'s mean energy is expected to decrease since the equilibrium with matter
is now preserved out to regions where the fluid is more than 1 order of magnitude less dense and $\unit[{\sim} 3]{MeV}$ colder.

\subsection{Impact of weak magnetism, RMF effects, and nucleon decays}
\label{sec:energy_shift}

\begin{figure*}
    \centering
    \includegraphics[width = \textwidth]{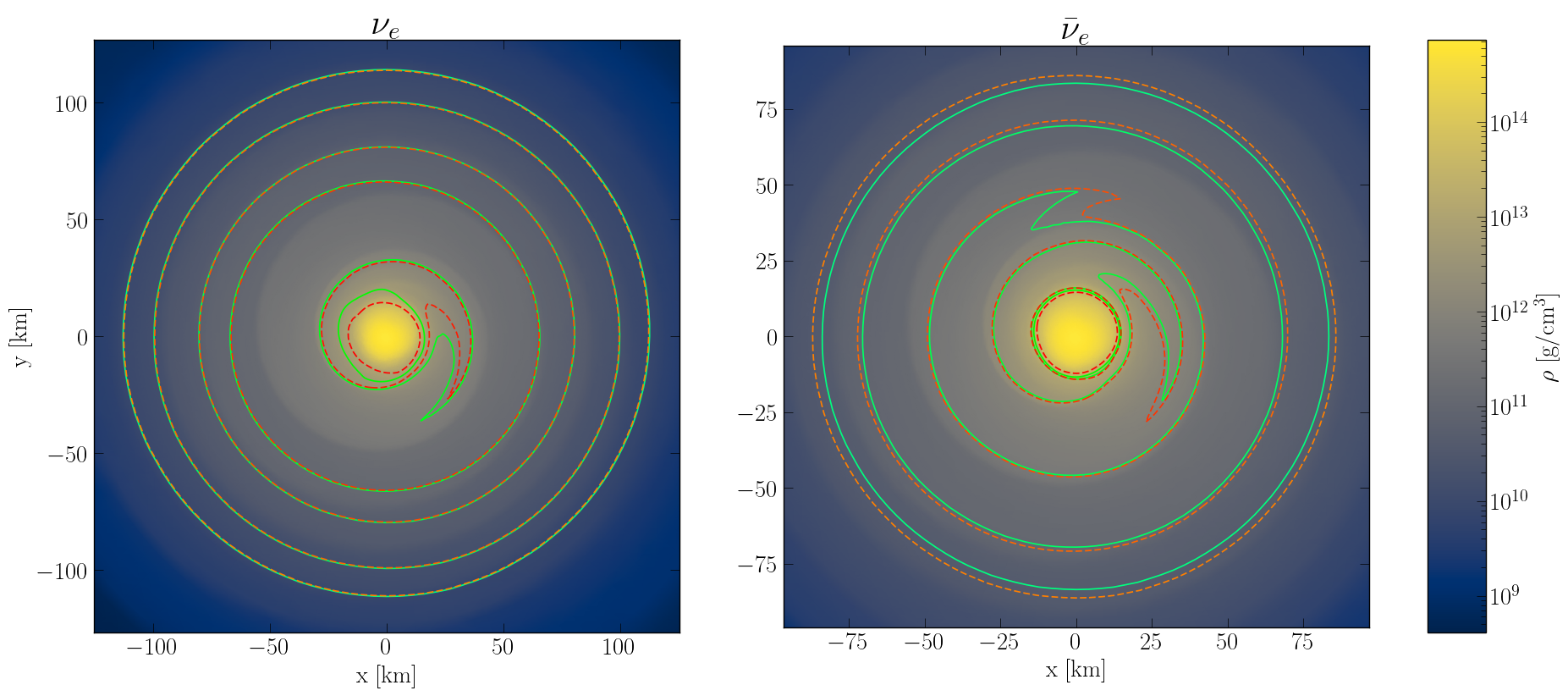}
    \caption{Energy-dependent, equilibration neutrino surfaces on the $xy$ plane, obtained with two different
    treatments for the opacities of $\beta$ processes. Dashed red lines refer to the case in which only bare
    inverse mean free paths for absorptions on nucleons are employed, while solid green ones to the case in which
    weak magnetism, \ac{RMF} effects and inverse nucleon decays are accounted for.
    Color coded is the rest mass density on the plane.
    The left (right) panel refers to electron (anti)neutrinos.
    Different color saturations are used to distinguish the neutrino surfaces of the following energies:
    3, 5, 7.5, 10, 17.5, and 25 $\unit[]{MeV}$, moving from inner to outer regions.
    Note the different spatial resolutions in the two panels due to the different typical size of the neutrino
    surfaces.}
    \label{fig:DD2_M12980-12980_M1_LR-40.02-xy_plane-all_reactions-m1_distr-eqOpcts_True-anue-CorrsCompareWDashed}
\end{figure*}

The red and purple dot-dashed lines in Figs.~
\ref{fig:DD2_M12980-12980_M1_LR_t_12390.4_n_50_rho_7.0e+14} and
\ref{fig:DD2_M12980-12980_M1_LR_t_12390.4_n_50_rho_1.0e+14} represent the
neutrino spectral emissivity and inverse mean free path due to $\beta$ processes
when no microphysics effects are included.
The direct comparison with the red and purple solid lines in the same panels highlights the 
importance of weak magnetism and \ac{RMF} effects. At the same time, the comparison between red and
purple lines addresses the importance of (inverse) nucleon decays and their dependence on the
inclusion of microphysics effects.

While $j_{\beta,\nue}$ is insensitive to the inclusion of microphysics effects for $k \gtrsim \mu_e$
(roughly corresponding to the $\nue$ mean energy in the case of a very degenerate electron gas, \ie, 
$\sim T F_5(\eta_e)/F_4(\eta_e) \sim (6/5) \mu_e$), below that threshold it is significantly enhanced (up to
4 orders of magnitude), mostly by \ac{RMF} effects.
Because of detailed balance, also the inverse mean free path is
increased by a similar amount below the electron Fermi energy. On the other hand, weak magnetism slightly
suppresses $\lambda_{\beta,\nue}^{-1}$ above ${\sim} \unit[200]{MeV}$, independently on the thermodynamic conditions.
In the case of electron antineutrinos, the emissivity is marginally increased (up to a factor of a few,
note the many orders of magnitude spanned by the vertical axes)
by the inclusion of \ac{RMF} effects (compensating for the attenuation induced by the weak
magnetism) for antineutrino energies above the mean
production energy by positron captures,
\ie, $k \gtrsim T F_5(- \eta_e)/F_4(-\eta_e) \sim 5 T$.
Below that value, the presence of $\Delta U > 0$ kinematically suppresses positron captures
(see \secref{reaction_comparison}), but the additional emissivity contribution from neutron decays
largely compensates for it, especially for soft antineutrinos. Concerning $\lambda_{\beta,\anue}^{-1}$,
we again observe a partial suppression at high antineutrino energies
due to weak magnetism and a significant enhancement
at low energies induced by the inclusion of inverse neutron decays.
We conclude that, overall, weak magnetism, \ac{RMF} effects, and (inverse) nucleon decays have a
significant impact on $\beta$ processes, possibly affecting
the diffusion and the equilibration timescales.

We would like to estimate if these effects could also affect the spectrum of neutrinos streaming at a
large distance from the remnant.
Using an approach similar to the one presented in \secref{NES_NPS}, in
\figref{DD2_M12980-12980_M1_LR-40.02-xy_plane-all_reactions-m1_distr-eqOpcts_True-anue-CorrsCompareWDashed}
we compare the
energy-dependent, equilibrated neutrino surfaces for the decoupling of electron-type (anti)neutrinos
obtained by considering two different
treatments for the opacities for $\beta$ processes. The dashed red lines refer to the case in which
only bare inverse mean free paths for absorptions on nucleons are employed, as presented in
Ref.~\cite{Bruenn:1985en}, while solid green ones to the case in which weak magnetism, \ac{RMF} effects, and
inverse nucleon decays are accounted for.
The left panel in
\figref{DD2_M12980-12980_M1_LR-40.02-xy_plane-all_reactions-m1_distr-eqOpcts_True-anue-CorrsCompareWDashed}
shows that only the electron neutrino surfaces at lower energies, \ie, $3$ and $\unit[5]{MeV}$,
are affected when considering
an improved description of $\beta$ processes.
The regions enclosed by these two surfaces are enlarged
following the enhancement of $\nue \, n$ absorptions for neutrinos
below the $\nue$ mean energy. This effect is entirely ascribable to
the inclusion of $\Delta U$, which ranges between $\unit[{\sim} 0.3]{MeV}$ and
$\unit[{\sim} 8]{MeV}$ in correspondence of the decoupling conditions (\cf
\figref{thermo_DD2_M12980-12980_M1_LR_axis_x_t_12380.8_rl_3}), since weak magnetism only impacts the opacities
of $\nue$ neutrinos with $k \gtrsim \unit[200]{MeV}$ and inverse proton decays
are kinematically suppressed everywhere. 
Neutrino surfaces of higher energy are not modified as $\Delta U$ becomes subdominant
to the bare nucleon mass difference when moving to outer and less dense regions.
The situation in the electron antineutrino case is more varied, as shown in the right panel of
\figref{DD2_M12980-12980_M1_LR-40.02-xy_plane-all_reactions-m1_distr-eqOpcts_True-anue-CorrsCompareWDashed},
since several of the included effects reveal to be significant, depending on the thermodynamic conditions.
Weak magnetism reduces $\lambda_{\beta, \anue}^{-1}$ everywhere,
but its importance increases with the antineutrino energy, coherently with the fact that
it includes the phase-space reduction due to the nucleon recoil as well.
Therefore, it affects particularly the outermost surfaces,
as high-energy antineutrinos decouple at lower matter densities. 
Indeed, we observe a shrinkage of
antineutrino surfaces with $k \unit[\geq 10]{MeV}$ that is directly 
connected to the inclusion of weak magnetism,
as they are located at densities $\unit[\lesssim 10^{12}]{g \, cm^{-3}}$,
where $\Delta U$ is negligible and inverse neutron decays consequently only absorb sub-MeV antineutrinos.
Moving to inner regions, \ac{RMF} effects become more relevant given the progressively
higher $\Delta U$ values, and combine with weak magnetism
in decreasing the opacity of $\anue \, p$ absorptions when $\rho \gtrsim \unit[10^{12}]{g \, cm^{-3}}$
(\cf Figs.~\ref{fig:DD2_M12980-12980_M1_LR_t_12390.4_n_50_rho_1.0e+13} and
\ref{fig:DD2_M12980-12980_M1_LR_t_12390.4_n_50_rho_1.0e+12}).
The marginal reduction in the size of $\anue$
surfaces at $7.5$ and $\unit[5]{MeV}$
is in fact a result of the interplay of the two effects.
Moving to lower antineutrino energies, decoupling at even higher densities,
the impact of weak magnetism becomes negligible while the nucleon interaction potential
difference at some point surpasses the energy of the particle experiencing last scattering with matter.
Recall that \ac{RMF} effects change the energy threshold of $\beta$ processes,
such that all electron
antineutrinos with $k \lesssim \Delta m^* + \Delta U$ are absorbed via inverse
neutron decays rather than interacting only with protons. 
As a result, inverse neutron decays push the innermost antineutrino surface ($k = \unit[3]{MeV}$) to slightly larger radii,
since they occur more frequently than $\anue \, p$ absorptions would in the absence of \ac{RMF} effects
(\cf \figref{DD2_M12980-12980_M1_LR_t_12390.4_n_50_rho_1.0e+14}).

\section{Results: energy integrated opacities}
\label{sec:gray}

\subsection{Reaction comparison}
\label{sec:r}

\begin{figure*}
    \centering
    \includegraphics[width=\textwidth]{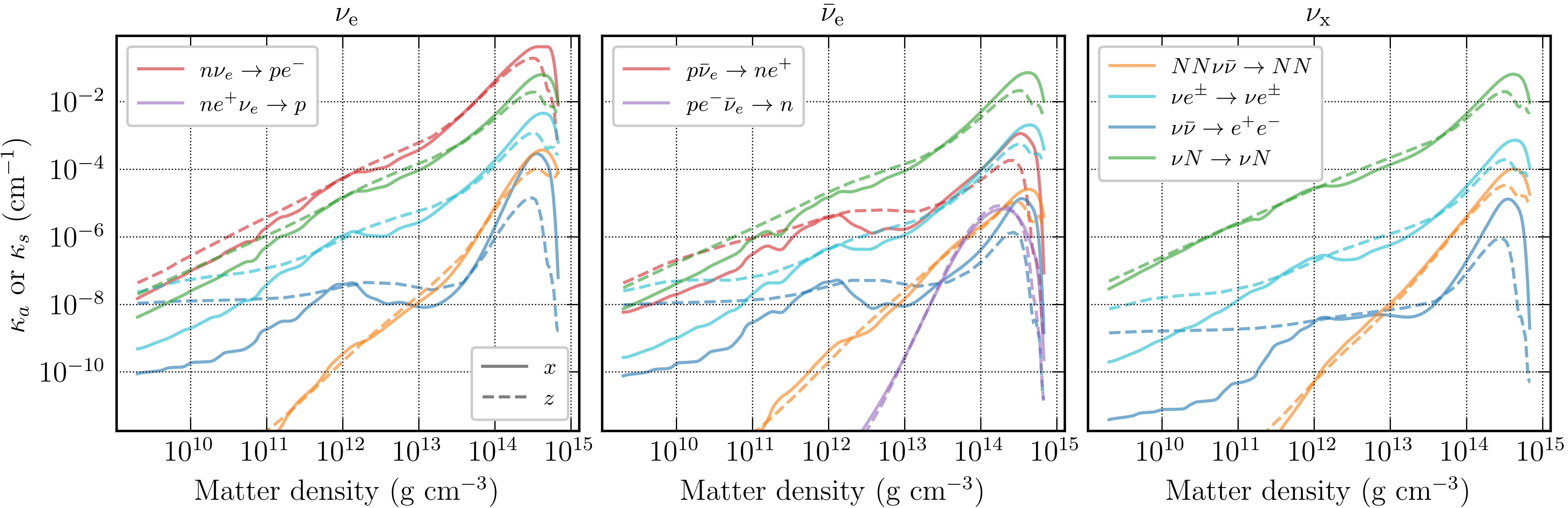}
    \caption{Gray absorption and scattering opacities
    as a function of the rest mass density along the positive $x$ (solid) and $z$ axis (dashed)
    of the numerical simulation under study. Different colors refer to the contribution of different
    reactions, while different panels to different neutrino flavors.}
    \label{fig:gray_x_vs_z_DD2_M12980-12980_M1_LR_12380.8_eos_CompOSE_n_50_kappa}
\end{figure*}

The discussion in \secref{spectral} focuses on the role of the different reactions in relation to
the neutrino energy. However, depending on the local thermodynamic and radiation conditions,
some energies could be more relevant than others.
Gray emissivities and opacities (see \secref{energy_integrated_rates}) are informative in this respect, as they account for the distributions of reacting particles.
Figure~\ref{fig:gray_x_vs_z_DD2_M12980-12980_M1_LR_12380.8_eos_CompOSE_n_50_kappa} compares the gray absorption and
scattering opacities for different reactions, computed according to Eqs.~
(\ref{eq:gray_abs_opacity_energy}) and (\ref{eq:iso_scatteringcoefficient}), for the profiles depicted in
\figref{thermo_DD2_M12980-12980_M1_LR_axis_x_t_12380.8_rl_3} (solid lines) and
\figref{thermo_DD2_M12980-12980_M1_LR_axis_z_t_12380.8_rl_6} (dashed lines), \ie,
along the positive $x$ and $z$ axis of the simulation under consideration.
In the very center of the remnant ($\rho \approx \unit[7 \times 10^{14}]{g \, cm^{-3}}$),
the main contribution to the opacity of electron neutrinos (left panel) comes from
isoenergetic scattering off nucleons and $\beta$ processes. In these conditions, the latter
are partially suppressed by the Pauli blocking of the dense electron gas.
At slightly lower densities ($\rho \sim \textrm{a few} ~ \unit[10^{14}]{g \, cm^{-3}}$), where the
temperatures are on average larger, \ie, $T \lesssim \unit[45]{MeV}$ ($\unit[20]{MeV}$) along the $x$ ($z$)
axis, \cf
\figref{thermo_DD2_M12980-12980_M1_LR_axis_x_t_12380.8_rl_3}
(\figref{thermo_DD2_M12980-12980_M1_LR_axis_z_t_12380.8_rl_6}), 
the efficiency of $\nue \, n$ absorptions
increases sharply since electrons are less degenerate.
As a consequence, when moving away from the center, the absorption opacity of $\nue$ neutrinos surpasses in
magnitude the scattering one.
The former is primarily contributed by $\beta$ processes also in the outer part of the profiles.
In fact, the integrated opacities for the other inelastic reactions seem to be subdominant everywhere,
consistently with the behavior of spectral emissivities and inverse mean free paths discussed in
\secref{reaction_comparison}.

Differently from the electron neutrino case, the opacity for electron antineutrinos (middle panel) is dominated by isoenegetic scattering
off nucleons for any condition along the profile, except for its very outer part ($\rho \lesssim {\rm a ~ few} ~ \unit[10^{10}]{g \, cm^{-3}}$) where its contribution becomes comparable to the one of other inelastic reactions.
Neutron degeneracy inhibits $\anue \, p$
absorptions at the highest densities, making scattering off $e^\pm$ the most efficient channel through which
electron antineutrinos exchange energy with matter deep inside the remnant.
The $\anue$ opacity associated with $\beta$ processes grows along the $x$ ($z$) axis in correspondence
of the region with higher temperatures, but it does not exceed the one
for $\anue \, e^\pm$ scattering until $\rho \lesssim \unit[10^{14}]{g \, cm^{-3}}$
($\unit[4 \times 10^{13}]{g \, cm^{-3}}$). Pauli blocking of neutrons in
the center also impacts the opacity associated with inverse neutron decays, whose shape closely follows the one of
$\Delta U$ as it determines how extended is the energy range over which this process is active
(see discussion in \secref{reaction_comparison}). Nonetheless, even where $\Delta U$ peaks, inverse neutron
decays are not as important as other reactions, since they only absorb (relatively) soft neutrinos which do not
weight much on the energy-averaged opacity.

Heavy-type neutrinos (right panel) primarily diffuse within the fluid through isoenergetic scattering off nucleons at any
density, as a result of the inability of charged-current processes. Among the inelastic reactions, they are mostly
subject to \ac{NEPS} given the conditions along the profiles.
Inverse NN bremsstrahlung partially contributes at high enough densities
($\rho \gtrsim \unit[10^{14}]{g \, cm^{-3}}$), but at larger radii its relevance rapidly fades away
due to the strong density dependence of the reaction and to its pair nature. We also notice how the conversion of
$\nu \bar{\nu}$ pairs into $e^+e^-$ is typically subdominant, for any flavor, down to ${\sim} \unit[10^{13}]{g \, cm^{-3}}$,
while it becomes more relevant than the inverse NN bremsstrahlung below that density, since it only depends
on the neutrino density and not also on the matter density.

According to
\figref{gray_x_vs_z_DD2_M12980-12980_M1_LR_12380.8_eos_CompOSE_n_50_kappa}, 
the magnitudes and the reaction hierarchy of the energy-integrated opacities are qualitatively in agreement when
comparing their dependence along the $x$ and $z$ axes. Nonetheless, we still observe some minor differences
that are common to different flavors and reactions. Opacities along the $x$ axis peak at higher values as a
result of the higher peak temperature of matter on the equatorial plane
(\cf Figs.~\ref{fig:thermo_DD2_M12980-12980_M1_LR_axis_x_t_12380.8_rl_3} and
\ref{fig:thermo_DD2_M12980-12980_M1_LR_axis_z_t_12380.8_rl_6}). On the other hand, opacities along the
$z$ axis exhibit a shallower decrease in optically-thin conditions,
\ie, for $\rho \lesssim \unit[10^{11} - 10^{12}]{g \, cm^{-3}}$.
This is a combined effect of the dependence of the opacity on the local neutrino density, whose dilution factor depends on the inverse squared distance from the central remnant, and of the sharper density decrease along the $z$ direction.
Therefore, for a given $\rho$, neutrinos are more abundant along the polar direction than on the equator,
as the given density corresponds to a smaller distance to the center.

\subsection{Impact of weak magnetism and RMF effects}
\label{sec:corrections}

\begin{figure}
    \centering
    \includegraphics[width=\columnwidth]{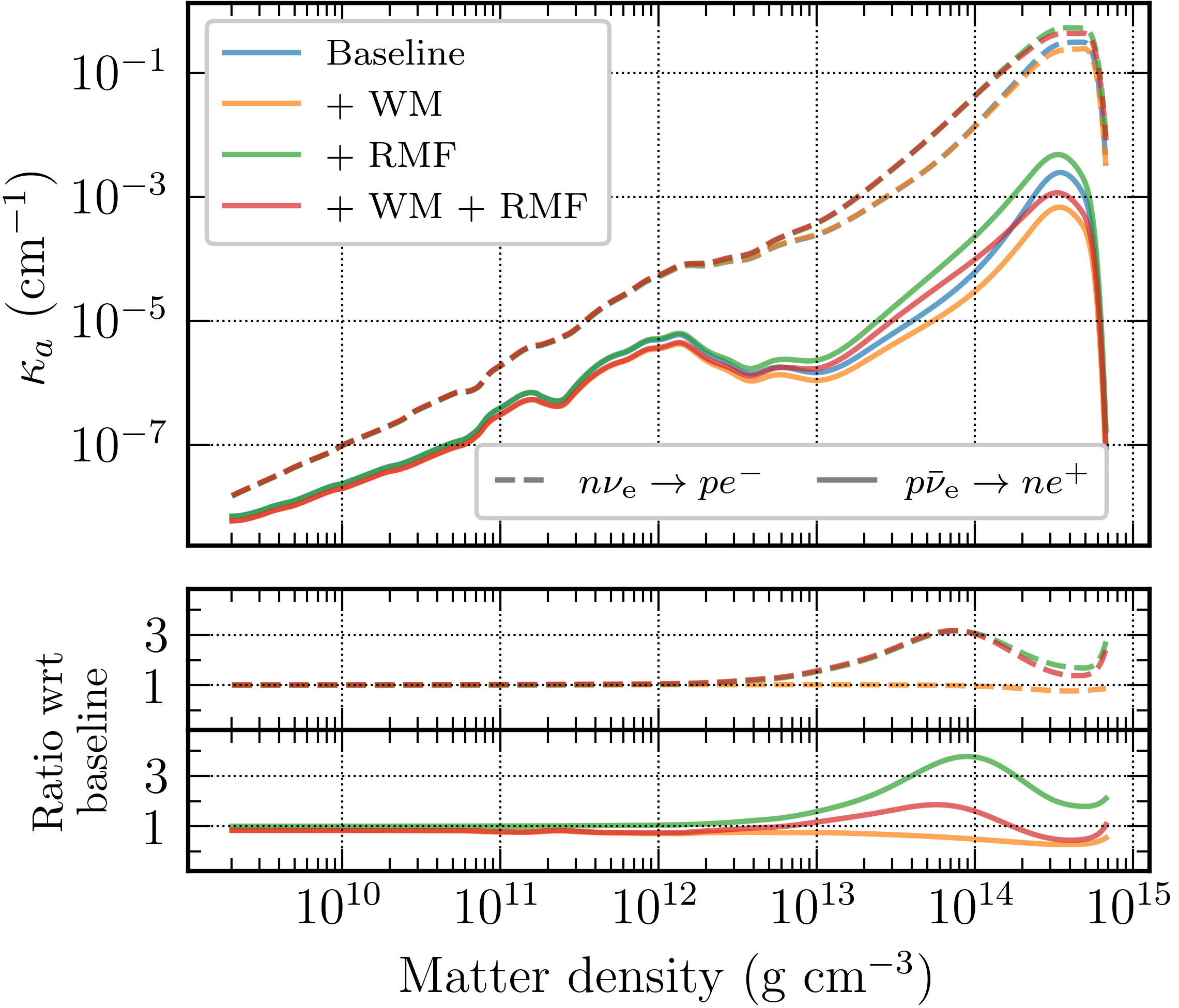}
    \caption{Top panels: impact of different effects on the gray absorption opacity associated with neutrino
    absorptions on nucleons as a function of the matter density along the positive $x$ axis.
    Middle and bottom panels: ratios with respect to the bare opacity.
    Dashed (solid) lines refer to $\nue \, n$ ($\anue \, p$) absorptions. The bare absorption opacity is shown in blue,
    the one accounting only for weak magnetism in orange, the one accounting only for \ac{RMF} effects in green,
    while the one including both of them in red.}
    \label{fig:gray_corrections_DD2_M12980-12980_M1_LR_axis_x_t_12380.8_rl_3_n_50_kappa_stim}
\end{figure}

Figure \ref{fig:gray_corrections_DD2_M12980-12980_M1_LR_axis_x_t_12380.8_rl_3_n_50_kappa_stim} exhibits the impact of
weak magnetism and \ac{RMF} effects on the gray absorption opacity associated with the
absorption of electron neutrinos (dashed curves) and antineutrinos (solid curves) on nucleons,
as a function of the matter density along the positive $x$ axis.
Inverse nucleon decays are affected by these effects too, but we do not consider them here as 
their contribution to the total $\kappa_{a,x}$ is subdominant, or even totally negligible, as shown in
\figref{gray_x_vs_z_DD2_M12980-12980_M1_LR_12380.8_eos_CompOSE_n_50_kappa}.
We observe that the opacity of electron neutrinos is mostly affected by the nucleon interaction potential
shift, $\Delta U$, which increases the efficiency of $\nue \, n$ absorptions up to a factor of ${\sim} 3$ in the
inner regions ($\rho \gtrsim \unit[10^{13}]{g \, cm^{-3}}$). 
On the other hand, the impact of weak magnetism is only marginal and limited to supranuclear densities.
In the case of electron antineutrinos, the two effects compete
with each other for high enough densities. \ac{RMF} effects can increase
the magnitude of the spectral stimulated $\anue$ absorptivity up to a factor
of a few at relevant energies (see \secref{energy_shift}),
but, at the same time, $\Delta U$ shifts the kinematic energy threshold of the reaction, increasing the lower
limit on the energy of antineutrinos that can be absorbed on protons.
Therefore, in the regions where \ac{RMF} effects are relevant, the enhancement is partially weakened once we weight
the spectral integrand by the antineutrino
occupation number when computing the energy-integrated opacity.
This is reflected by the fact that the \ac{RMF}-over-bare opacity ratio does not follow exactly the shape
of the $\Delta U$ distribution along the profile
(\cf \figref{thermo_DD2_M12980-12980_M1_LR_axis_x_t_12380.8_rl_3}).
It peaks instead at slightly lower densities than $\Delta U$, where the influence of $\anue \, p$
absorptions extends over a wider energy range.
Conversely, the ratio is lower than one in the case of weak magnetism because it decreases the magnitude of
the spectral stimulated $\anue$ absorptivity, with a reduction that is more and more significant as
antineutrinos are more energetic.
Therefore, the impact on the integrated opacity becomes milder when moving to larger radii, because the fraction of
high-energy antineutrinos becomes progressively less significant.
However, differently from \ac{RMF} effects, the impact due to weak magnetism survives down
to $\rho \unit[{\sim} 10^{10}-10^{11}]{g \, cm^{-3}}$, affecting the typical energy of
electron antineutrinos that are emitted from the system (see \secref{energy_shift}).
 
\subsection{Estimating the optically thin opacities vs reconstructing the neutrino
distribution function}
\label{sec:E2_vs_nu_f}

\begin{figure*}
    \centering
    \includegraphics[width=\textwidth]{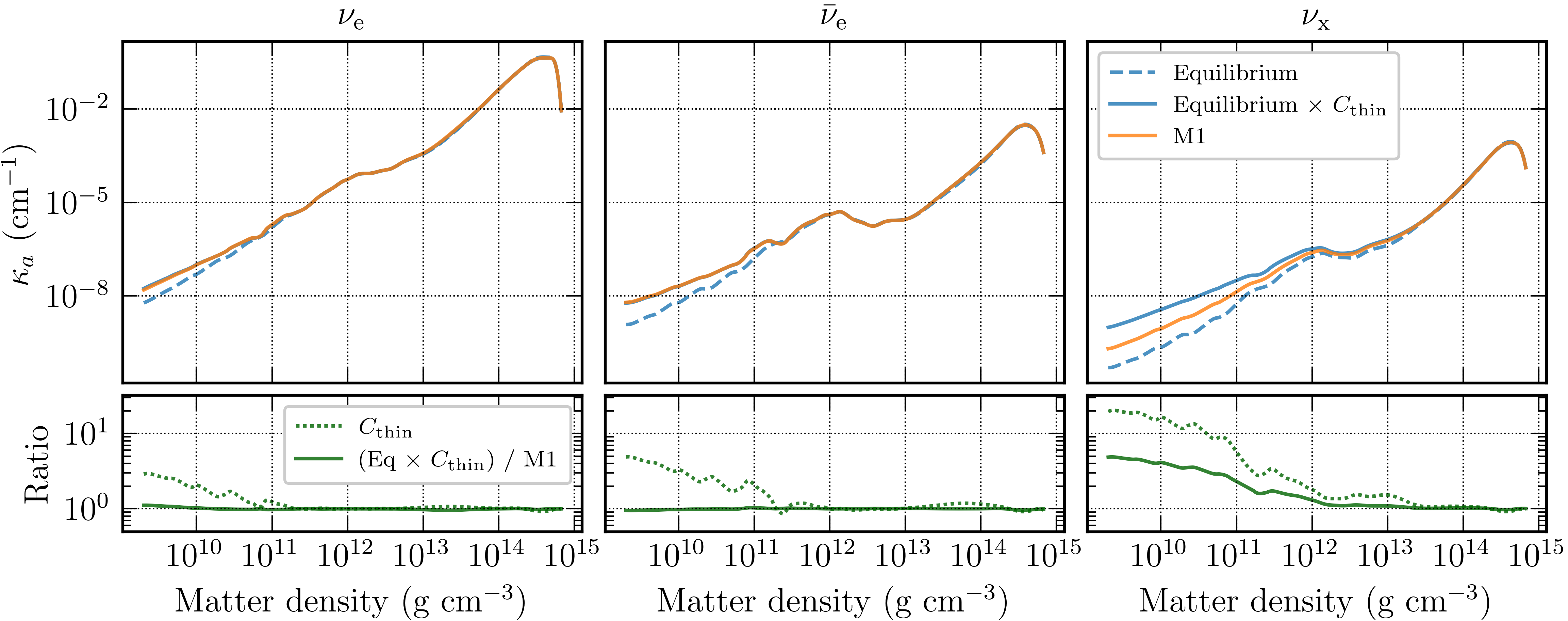}
    \caption{Top panels: total gray absorption opacities along the positive $x$ axis
    obtained reconstructing $f_\nu$ from
    \mone radiation quantities (solid orange) or assuming equilibrium $f_\nu$ (dashed blue),
    eventually corrected for thin conditions (solid blue) as in \eqref{opt_thin_corr_factor}.
    Bottom panels: ratio between the solid curves in the top panels (solid green) and
    correction factor applied to equilibrium opacities, $C_{\tn{thin}}$ (dotted green), corresponding to the ratio
    between the blue curves in the top panels.
    Different panels refer to different neutrino species.}
    \label{fig:gray_m1_vs_eq_DD2_M12980-12980_M1_LR_axis_x_t_12380.8_rl_3_eos_CompOSE_n_50_kappa_a_stim}
\end{figure*}

Gray neutrino transports require the knowledge of neutrino distribution functions for the evaluation
of energy-integrated emissivities and opacities. However, many approximated schemes that are nowadays employed
in simulations do not contain the full information about $f_\nu$. 
One common strategy is to assume equilibrium distributions at the fluid temperature and composition.
As this assumption is not well justified in optically thin conditions,
the computation of the opacities could be affected by some systematics,
especially in the case of $\nu\bar{\nu}$ processes due to their nonlinear dependencies on the 
distribution functions. In order to account for that,
the gray absorption and scattering opacities are usually corrected in the way
presented in \appref{nu_distribution}.

We discuss hereafter how opacities are affected by the specific assumptions made on the distribution
function. In this respect,
\figref{gray_m1_vs_eq_DD2_M12980-12980_M1_LR_axis_x_t_12380.8_rl_3_eos_CompOSE_n_50_kappa_a_stim} compares
the total absorption opacity along the positive $x$ axis 
in the case of $f_\nu$ reconstructed from local \mone neutrino quantities or assuming equilibrium
conditions (see \secref{nu_distribution}), eventually corrected by \eqref{opt_thin_corr_factor}.
As expected, from a given density on, the two approaches are no longer in agreement. This 
corresponds to the point where neutrinos decouple from the matter, which occurs around 
$\rho \unit[{\sim} 2 \times 10^{11}]{g \, cm^{-3}}$ for electron neutrinos, 
$\rho \unit[{\sim} 4 \times 10^{11}]{g \, cm^{-3}}$ for electron antineutrinos and
$\rho \unit[{\sim} 2 \times 10^{13}]{g \, cm^{-3}}$ for the other neutrino flavors. Nonetheless,
in the case of $\nue$ and $\anue$, the correction introduced on the equilibrium opacities
is still able to reproduce with a good accuracy the
ones reflecting the local radiation properties, even in optically thin conditions.
The correction for heavy-flavored neutrinos also starts to be effective once the decoupling happens, but it
overestimates the correct opacity by a factor of a few.
We can therefore conclude that using a blackbody distribution function, eventually correcting opacities
to account for the local temperature in optically thin regimes, works reasonably well when dealing with
electron (anti)neutrinos. In the case of heavy-flavored (anti)neutrinos, more attention should be paid.

\section{Performance of \bnsnurates}
\label{sec:performance}

\begin{figure}[t!]
  \centering
  \includegraphics[width=\columnwidth]{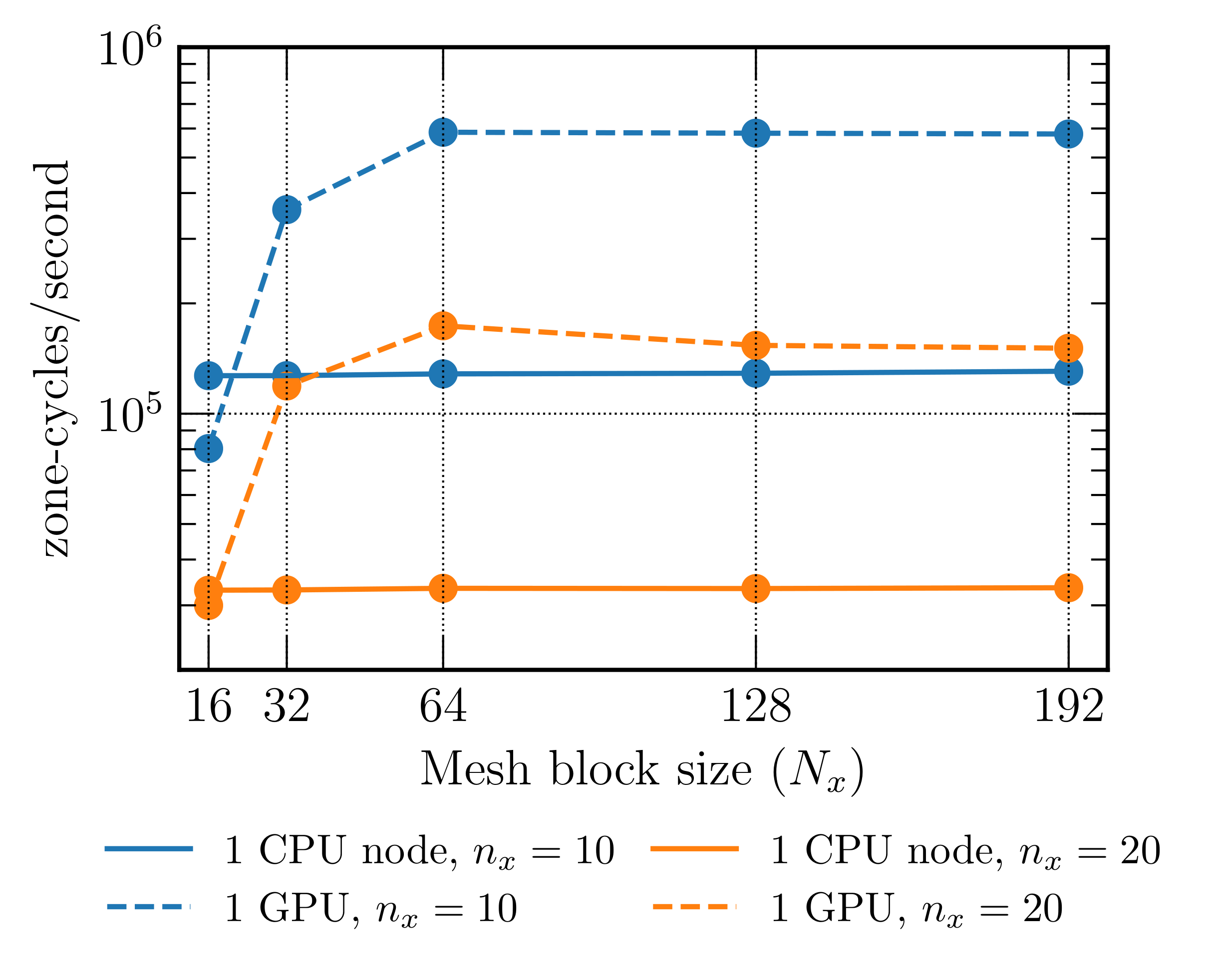}
  \caption{Central processing unit (CPU) and graphical processing unit (GPU)
  performance of \bnsnurates, measured in
    zone cycles per second as a function of domain size.
    For the comparison shown with blue (orange) lines we set $n = 10$ ($n = 20$)
    in the quadrature scheme.}
    \label{fig:performance_kokkos}
  \end{figure}

\bnsnurates is fully integrated with \textsc{Kokkos}, a C++ library that enables
single source performance portable codes \cite{Trott:2022tlg,Carter:2014cts}.
Thanks to this design, the library can run efficiently both on \acp{CPU} and
\acp{GPU}. In the latter case, the computation of spectral or energy-integrated
emissivities and opacities at different grid points can be offloaded to multiple \ac{GPU} cores.
We provide here an evaluation of the performance of \bnsnurates on \acp{CPU} and
accelerators, employing the supercomputing cluster \textsc{Perlmutter} at NERSC \cite{Perlmutter}.
For \ac{CPU} runs, the code
was compiled with \textsc{OpenMP} and was run on a single node with two AMD EPYC
7763 \acp{CPU}, totaling 128 cores. This is compared with performance on a
single Nvidia A100 \ac{GPU}. We ran a test problem in which the 
integrated (number and energy) emissivities and opacities for all reactions
are computed for thermodynamic conditions
chosen randomly from a \ac{CCSN} profile on three dimensional
grids with $16^3$, $32^3$, $64^3$, $128^3$, and $192^3$ points. We repeated the
test considering two different numbers of quadrature points for the numerical
integration routine, namely $n=10$ and $n=20$ (see \appref{quadrature_scheme}).

As can be seen in \figref{performance_kokkos}, for a small number of points
($N_x = 16^3$), the performance of \bnsnurates on the \ac{GPU} is worse than on
the \acp{CPU}. This is likely due to the intrinsic overhead 
associated with \textsc{Kokkos} in offloading operations on the \ac{GPU} cores
not being offset by the increased computation speed. For
increased computational loads, however, the 
code is significantly more
performing on a single \ac{GPU}, in this case by a factor of ${\sim} 5$. In the
physical scenario of \ac{BNS} merger simulations that we target, the typical
number of points in the computational domain can be very large, so that
\bnsnurates will operate in this latter regime. We are indeed planning to
integrate \bnsnurates with the \ac{GPU}-based hydrocode \textsc{AthenaK}
\cite{Fields:2024pob}.
Doubling the number of quadrature points on each integration axis causes
a performance hit by a factor of ${\sim} 3.5 - 3.8$ for $N_{x}>64$, \ie, consistent
with, but somewhat smaller than, the factor of $4$ that would be expected
given the quadratic scaling of the number of kernel evaluations with $n$
(see \appref{quadrature_scheme}).

\section{Discussion and conclusions}
\label{sec:conclusions}

In this work we have presented \bnsnurates, a novel open-source software library designed
for performance-portable computation of realistic neutrino-matter interaction rates,
with particular focus on regimes relevant to \ac{BNS} mergers. We have then
applied \bnsnurates to the evaluation of both spectral and energy-integrated
emissivities and opacities for different conditions extracted from a \ac{BNS} merger simulation evolved with \mone neutrino transport.

The aim of this study is twofold. On one hand we showcased and tested the
capabilities of this new tool. On the other, we provided a general and
comprehensive overview of the various reactions undergone by neutrinos of different species
for typical \ac{BNS} merger conditions, employing a higher degree of accuracy and
realism compared to the approaches commonly used in this field. We found that
some of the additions and improvements implemented into the library lead to
significant differences in the characterization of neutrino interactions in the regimes under consideration. This analysis
has highlighted some important aspects, which might stimulate discussion about
the current status of neutrino rate modeling in \ac{BNS} mergers.

We found that inelastic scattering off $\tn{e}^\pm$, an often neglected process
in \ac{BNS} mergers, affects the behavior of heavy-flavored neutrinos in
general, as well as the one of electron (anti)neutrinos in very high density regions
($\rho {\sim} \unit[7 \times 10^{14}]{g \, cm^{-3}}$),
where the degeneracy of neutrons and electrons suppresses the importance of
$\beta$ processes. In particular, the inclusion of $\nu \, e^\pm$ scattering
provides an additional significant contribution to the $\nux$ opacity that
pushes the heavy-type neutrino surfaces to outer radii,
softening their emission spectrum. This effect is similar to what is observed
in the cooling phase of protoneutron stars, see, \eg, \cite{Fischer:2009af,Hudepohl:2009tyy}.

We also showed that weak magnetism and \ac{RMF} effects can significantly alter $\nue$
and $\anue$ bare emissivities and opacities. \ac{RMF} effects play a relevant role for the dynamics
of both species in the regions where the nucleon interaction potential difference, $\Delta U$, is non-negligible, \ie, $\rho
\gtrsim \unit[10^{13}]{g \, cm^{-3}}$. Weak magnetism is instead more impactful for
electron antineutrinos, since, differently from the $\nue$ case, it is important also in outer, less
dense regions, affecting the typical conditions at which $\anue$ decouple, see, \eg, \cite{Horowitz:2001xf}.
We also include the
contribution of (inverse) neutron decays, which we found to be particularly
relevant for soft $\anue$. Their importance is coupled with the one of \ac{RMF}
effects, as the upper limit on the energy of antineutrinos they can emit (absorb)
is established by the value of $\Delta U$, in agreement with previous findings, see, \eg, \cite{Oertel:2020pcg}.

\bnsnurates attempts to go beyond the assumption of \ac{LTE} conditions for
neutrinos as it exploits gray radiation moments to reconstruct the distribution
functions. We tested how the two approaches compare with each other and impact
on the interactions. As is to be
expected, they differ significantly once the decoupling from matter occurs, but
in the case of electron (anti)neutrinos, gray opacities in
optically thin conditions can be still correctly estimated even assuming equilibrium,
thanks to the application of a correction factor that accounts for the local
temperature of the radiation. On the other hand, the opacities of heavy-flavored
(anti)neutrinos are still biased even after the introduction of this correction,
meaning that a more realistic estimate of the neutrino distribution such as
the one we implement is required to avoid the introduction of systematic
errors in the opacities.

Despite being informative about the importance of the different neutrino
processes in \ac{BNS} mergers, the present study is still subject to some
caveats and limitations. For example, the implemented inelastic reactions do not have any
angular dependence, to avoid to perform additional numerical integrations when
computing gray emissivities and opacities.
In fact, we truncate the Legendre expansion of the reaction kernels at
the monopole term (dipole term for the isoenergetic scattering off nucleons). This
choice is motivated by the necessity of evaluating neutrino interactions quickly enough to
allow \textit{in situ} coupling of \bnsnurates to hydrodynamic simulation codes.
The library realism could be further increased by considering more terms of the
Legendre expansion and by extending the reconstruction of the distribution
functions to the angular part, exploiting the additional information about the
neutrino flux density.

Furthermore, individual reactions could benefit from an additionally improved treatment. As
an example, it is important to assess if the description of the NN bremsstrahlung is reliable enough,
as we found that it is the most relevant reaction at very high
densities ($\rho {\sim} \unit[7\times 10^{14}]{g \, cm^{-3}}$) for soft neutrinos of
all flavors. Notice that this is valid despite modeling this process following Ref.~\cite{Hannestad:1997gc},
which predicts considerably smaller emissivities and opacities at high densities with respect to the more widely
adopted, yet less detailed, prescription of Ref.~\cite{Burrows:2004vq} (see also
\cite{Betranhandy:2020cdf} for a comparison in the context of \acp{CCSN}).
Nonetheless, both approaches rely on a one-pion exchange approximation, which is not appropriate at high densities ($\rho \gtrsim \unit[10^{14}]{g \, cm^{-3}}$).
Modified Urca rates, at present absent from our implementation, may also be relevant to accurately
describe the precise neutrino dynamics in the central regions of the merger
remnant (see, \eg, \cite{Most:2022yhe,Espino:2023dei}). 
An accurate and complete implementation of charged-current reactions is essential to study the dynamical equilibration
of the proton fraction via weak interactions 
and the possible appearance of bulk viscosity, given its potential to alter the postmerger
dynamics and its observables; see, \eg, \cite{Alford:2018lhf,Radice:2021jtw,Hammond:2021vtv,Most:2021ktk,Espino:2023dei}.
We leave such improvements to future work, but we wish to stress that
our publicly available library is going to be actively and continuously maintained and improved. In
fact, work on tackling some of the issues mentioned has already started.

Finally, the postprocessing approach we have adopted in this work is of course
not fully consistent and cannot capture the dynamical impact of the detailed
interactions on the \ac{BNS} merger system. To do so, \bnsnurates should be directly coupled to the
neutrino transport scheme employed in a dynamical simulation, evolving the
system consistently starting from premerger conditions. Work on achieving this goal is also
under way.

\begin{acknowledgments}

F.M.G. and partially also A.P. are supported by the European Union under NextGenerationEU, PRIN 2022
Project No.~2022KX2Z3B. M.B. and D.R. are supported by
the U.S.~Department of Energy, Office of Science, Division of Nuclear Physics
under Awards No.~DE-SC0021177 and No.~DE-SC0024388.
D.R. also acknowledges support from the Sloan Foundation and from the National Science Foundation under Grants No.~PHY-2020275, No.~AST-2108467, No.~PHY-2116686, and No.~PHY-2407681.
We acknowledge the EuroHPC Joint Undertaking for awarding this project access to
the EuroHPC supercomputer LUMI, hosted by CSC (Finland) and the LUMI consortium
through a EuroHPC Extreme Scale Access call (EHPC-EXT-2022E01-046).
This research used resources of the National Energy Research Scientific Computing
Center, which is supported by the Office of Science of the U.S. Department of Energy under
Contract No. DE-AC02-05CH11231.

\end{acknowledgments}

\section*{Data Availability}

Part of the data presented in this manuscript are publicly
available \cite{Perego_BNS_NURATES_2025}. The remaining data supporting the findings
of this study will be made available upon reasonable
request to the corresponding author.

\appendix

\section{REACTION IMPLEMENTATIONS}
\label{sec:reactions_implementation}

We report in the following the explicit expressions of the spectral emissivity (for $\beta$ processes) and
kernels (for other reactions), as implemented
in \bnsnurates. Inverse mean free path and kernels for inverse reactions are consistently computed exploiting the relations reported in \secref{neutrino_reactions}. Throughout this section $G_F\simeq\unit[8.96\times10^{44}]{MeV \, cm^3}$ is the Fermi constant, $\hbar$ is
the reduced Planck constant, $c$ is the speed of light, $m_{\tn{e}}$ is the
electron mass, $f_{\tn{e}^\pm}$ is the Fermi-Dirac distribution function for
$\tn{e}^\pm$, $Q \equiv \Delta m =  m_\n - m_\p \simeq \unit[1.29]{MeV}$ is the bare nucleon
mass difference, $\theta_w \simeq 28.2^\circ$ is the Weinberg angle, and $g_V=1$ and
$g_A=1.23$ are the nucleon form factors in the zero momentum
transfer limit.

\paragraph{$\beta$ processes.}
Following Ref.~\cite{Bruenn:1985en}, the spectral emissivity
for electron
(positron) captures reads as follows:
\begin{widetext}
  \begin{align}
    j_{\beta,\nue}(k) &=
                   \frac{G_F^2\,c}{\pi\,(\hbar c)^4}
                   \,\eta_{\p\n}\, (g_V^2+3g_A^2)\,(k+Q)^2
                   \sqrt{1-\frac{m_{\tn{e}}^2}{(k+Q)^2}}\,f_{\em}(k+Q)
                   \, \theta(k+Q-m_{\tn{e}}) \, W(k)\,,
                   \label{eq:spectral_emissivity_e_capture} \\
    j_{\beta,\anue}(k) &=
                         \frac{G_F^2\,c}{\pi\,(\hbar c)^4}
                         \,\eta_{\n\p}\, (g_V^2+3g_A^2)\,(k-Q)^2
                         \sqrt{1-\frac{m_{\tn{e}}^2}{(k-Q)^2}}\,f_{\ep}(k-Q)
                         \, \theta(k-Q-m_{\tn{e}}) \, \bar{W}(k)\,,
                         \label{eq:spectral_emissivity_p_capture}
  \end{align}
\end{widetext}
where $\theta$ is the Heaviside step function, which accounts for the reaction
kinematics. As we do not account for $\mu^\pm$ or $\tau^\pm$ leptons in the
system, $j_{\nux}(k)$ and $j_{\anux}(k)$ are identically equal to zero $\forall k$. The functions $W(k)$ and $\bar{W}(k)$
are the energy-dependent corrections due to the sum of phase-space, recoil, and
weak magnetism effects and are included following Ref.~\cite{Horowitz:2001xf}.
The quantities $\eta_{\p\n}$ and $\eta_{\n\p}$ account for the
nucleon final state blocking:
\begin{equation}
  \eta_{\p\n} = \frac{n_\n - n_\p}
  {\exp\left\{\left(\mu_\p-\mu_\n-Q\right)/T\right\}-1} \,,
  \label{eq:eta_pn}
\end{equation}
where $n_\p$ ($n_\n$) is the proton (neutron)
number density and $\mu_\p$ ($\mu_\n$) is the proton (neutron) relativistic
chemical potential. $\eta_{\n\p}$ can be obtained via the replacement $n
\leftrightarrow p$ in \eqref{eta_pn}. We also account for in-medium effects
associated to nucleon interactions following
Refs.~\cite{Hempel:2014ssa,Oertel:2020pcg}. Such effects can be implemented by
replacing $Q$ in Eqs.~(\ref{eq:spectral_emissivity_e_capture})-(\ref{eq:eta_pn}) with the nucleon energy
difference within the \ac{RMF} framework, \ie, $Q^* \equiv \Delta m^* + \Delta
U$. Neutrino spectral emissivities resulting from nucleon decays are related to
the ones of lepton captures through the crossing symmetry of the captured lepton
\cite{Oertel:2020pcg}. In fact, one can recover $j_{\nue}$ and $j_{\anue}$ for
nucleon decays through the substitutions $k\rightarrow-k$,
$Q^{(*)}\rightarrow-Q^{(*)}$, and $f_{\tn{e}^\pm}\rightarrow1-f_{\tn{e}^\mp}$ in Eqs.~
(\ref{eq:spectral_emissivity_e_capture}) and (\ref{eq:spectral_emissivity_p_capture}),
respectively.

\paragraph{Pair processes.}
For electron-positron annihilations (EPA), the zeroth-order term of the
Legendre expansion of the production kernels is modeled as
(see \cite{Pons:1998st})
\begin{equation}
  \begin{split}
    R^{\tn{pro},0}_{\tn{EPA},x}(k,k') & =
                              \frac{1}{2\pi}
                              \frac{G_F^2}{\hbar\,(\hbar c)^3}
                              \frac{T^2}{1-e^{(y+z)}} \times \\
                            & \big[
                              \alpha_{1,x} \Psi_0(y,z) +  \alpha_{2,x} \Psi_0(z,y)
                              \big] \, ,
  \end{split}
\end{equation}
where $y \equiv k/T$ and $z \equiv k'/T$ are the dimensionless energies of the
(anti)neutrino with flavor $x$ and of its corresponding antiparticle,
respectively. The coefficients $\alpha_{1,2}$ are defined as $\alpha_{1,\nue} = 1 + 2
\sin^2\theta_w$, $\alpha_{1,\nux} = -1 + 2 \sin^2\theta_w$, and $\alpha_{2,\nue} = \alpha_{2,\nux} = 2
\sin^2\theta_w$. For antineutrinos, the
coefficients are obtained by exchanging $\alpha_{1,x} \leftrightarrow \alpha_{2,x}$ of the corresponding neutrino. The
$\Psi_0$ function, whose explicit form is reported by Eq.~(11) in Ref.~\cite{Pons:1998st}, is
given by a linear combination of incomplete Fermi-Dirac integrals. We exploit
recursive formulas to express them in terms of the complete ones, which we
evaluate as described in Ref.~\cite{Fukushima:2015a}.

In the case of NN bremsstrahlung channels, the annihilation kernels do not depend on
the neutrino species and the monopole term of their Legendre expansion reads ($C_A=-1.26/2$)
\begin{equation}
  R_{\tn{Brem}}^{\tn{ann},0}(k, k') =
  3 \, C_A^2 \, \frac{G_F^2}{\hbar} \, n_{\tn{N}}
  S_\sigma(n_{\tn{N}};k + k') \, .
  \label{eq:brem_ann_kernel_0}
\end{equation}
For the $S_\sigma$ function we adopt the fitting formula provided by
Ref.~\cite{Hannestad:1997gc}, which is based on a one-pion exchange description of the
nucleon-nucleon interaction. The quantity $n_{\tn{N}}$ appearing in
\eqref{brem_ann_kernel_0} is the nucleon number density for the corresponding
bremsstrahlung channel ($n_\n$, $n_\p$, $\sqrt{n_\n\,n_\p}$ for neutron-neutron,
proton-proton, and neutron-proton, respectively). The total kernel is obtained by
summing \eqref{brem_ann_kernel_0} over the three channels. We also account for
the impact of medium modifications on the process by applying the
density-dependent correction factor defined in Ref.~\cite{Fischer:2016boc}, which
significantly decreases the magnitude of the bare kernel at high densities
($\gtrsim 70\%$ reduction for $\rho \gtrsim \unit[10^{14}]{g \, cm^{-3}}$).

\paragraph{Scattering processes.}
The first two Legendre coefficients of the isoenergetic scattering kernel expansion are defined in
Ref.~\cite{Bruenn:1985en} as the following, depending on the target particle
$N\in\{\n,\p\}$:
\begin{align}
  \left[{R^0_{\tn{iso}}(k,k') \atop R^1_{\tn{iso}}(k,k')}\right] =&
  2\pi\frac{G_F^2}{\hbar} \, \eta_{\tn{NN}} \, \delta(k-k') \times\\\nonumber
  &\left\{{\left[(h_V^N)^2+3(h_A^N)^2\right]W^N_0(k)\atop \left[(h_V^N)^2-(h_A^N)^2\right]W^N_1(k)}\right\} \, .
  \label{eq:iso_kernel}
\end{align}
The values of the coupling constants $h_V^N$ and $h_A^N$ are reported in
Ref.~\cite{Bruenn:1985en} and do not depend on the neutrino species, so the kernel expressions are the same for each $x$. The quantity $\eta_{\tn{NN}}$ results from the integration
over the nucleon phase-space variables and is computed considering a linear
interpolation between the results for the no-degenerate and degenerate limits for nucleons. The functions $W^N_0(k)$ and $W^N_1(k)$ are included to account for phase-space,
recoil, and weak magnetism effects associated with neutral-current reactions for
the zeroth and first coefficient, respectively. Such functions have been
calculated by expanding Eq.~(12) in Ref.~\cite{Horowitz:2001xf} to first order in $\omega$,
assuming $k=k'$.

In the case of inelastic neutrino-electron (NES) and neutrino-positron (NPS) scattering,
the isotropic term in the kernel expansion is computed following the approach in
Refs.~\cite{Bruenn:1985en,Mezzacappa:1993mab}, but performing the integration over the
electron energy analytically:
\begin{widetext}
  \begin{equation}
    \begin{split}
      R^{\tn{out},0}_{\genfrac(){0pt}{3}{\tn{NES}}{\tn{NPS}},x}(k,k') =
      &-\frac{1}{3\pi}\left[\frac{G_F}{(\hbar c)^3}\right]^2 (\hbar c)^2
        \,T^2\times\left\{ [1-\exp(z-y)]y^2 z^2 \right\}^{-1} \times \\
      &\bigg\{\frac{ \alpha_{1,x}^2+ \alpha_{2,x}^2}{5}
        \left[\tn{sign}(y-z)A^i_5+D^i_5 \right] +  \alpha_{1,x}^2
        (y + z) \left[ D^i_4 + 2 (y + z) D^i_3 + 6 y z D^i_2 \right] + \\
      &\hphantom{\bigg\{} 6  \alpha_{1,x}^2 y^2 z^2 D^i_1
        - {\rm sign}(y - z) [ ( \alpha_{1,x}^2 M_{y, z} -  \alpha_{2,x}^2 m_{y, z}) A^i_4 +
        2( \alpha_{1,x}^2 M_{y, z}^2 +  \alpha_{2,x}^2 m_{y, z}^2) A^i_3 ] \bigg\} \, .
    \end{split}
  \end{equation}
\end{widetext}
Here, $M_{y, z} \equiv \max(y, z)$ and $m_{y, z} \equiv \min(y, z)$ indicate the
maximum and minimum, respectively, between $y=k/T$ and $z=k'/T$. We also
introduce the functions $D^{i}_n=D^i_n(y, z) \equiv F_n(\eta_i - z) - F_n(\eta_i
- y)$ and $A^{i}_n=A^i_n(y, z) \equiv F_n(\eta_i - |y - z|) - F_n(\eta_i)$ to
express subtraction between complete Fermi-Dirac integrals of order $n$, where
$\eta_i$ is the degeneracy parameter of $i=\em$ ($i=\ep$) in the case of NES (NPS).
The total \ac{NEPS} \textit{out} kernel is obtained by summing $R_{\tn{NES},x}^{\tn{out},0}$ and
$R_{\tn{NPS},x}^{\tn{out},0}$.

\section{GAUSSIAN QUADRATURE SCHEME}
\label{sec:quadrature_scheme}

As discussed in \secref{energy_integrated_rates}, the evaluation of gray \mone
emissivities and opacities requires one- and two-dimensional integrations from zero
to infinite neutrino energies. In order to perform the integration, we split the
domain into two intervals, one ranging from $0$ to $t_{l,x}$ and the other from
$t_{l,x}$ to $+\infty$, with the splitting energy,
$t_{l,x}$, chosen depending on
the reaction $l$ and the neutrino species $x$. We map each interval onto the
$\left[ 0, 1 \right]$ segment by a change of variable. The contribution to the
total integral of each segment is evaluated via a Gauss-Legendre quadrature scheme with
$n$ points. Let $\{x_j\}$ and $\{w_j\}$ with $j=1,2,\dots,n$ be the
(dimensionless) quadrature nodes and weights in the $\left[ 0, 1 \right]$
interval, respectively. The integrands are then evaluated at the energies $k_j =
t_{l,x} \, x_j$ ($k_j = t_{l,x} / x_j$) for the first (second) segment.
Therefore, the quadrature of a generic energy-dependent integrand $G_{l,x}(k)$
reads
\begin{equation}
    \int \dd k \, G(k) = t \sum_{j=1}^n w_j \left[ G\left(t \, x_j\right)
    + \frac{G\left(t / x_j\right)}{x_j^2} \right]\,,
  \label{eq:1d_gauss_legendre_integration}
\end{equation}
where we dropped the $\{l,x\}$ dependence in the interest of readability. In this way, the
total number of evaluations of $G$ in \eqref{1d_gauss_legendre_integration} is
equal to $2n$, with $n$ set by the user at run-time. The results presented in the
manuscript have been obtained by setting $n=50$, unless
otherwise specified.

The contribution by $\beta$ processes to the total emissivity and absorption opacity of $\nue$ ($\anue$) is evaluated by splitting 1D integrals at the mean energy of (anti)neutrinos emitted
via electron (positron) captures \cite{Rosswog:2003rv}:
\begin{equation}
  t_{\nue} = T \, \frac{F_5(\eta_\tn{e})}{F_4(\eta_\tn{e})}\,,
  \quad
    t_{\anue} = T \, \frac{F_5(-\eta_\tn{e})}{F_4(-\eta_\tn{e})}\,.
\end{equation}
Instead, for the gray scattering opacity we split the domain at the local neutrino mean energy,
given by the ratio between the zeroth neutrino moments, \ie, $t_x =J_x/n_x$.

Equation~(\ref{eq:1d_gauss_legendre_integration}) can be easily generalized to treat
two-dimensional integration for reactions involving $\nu\Bar{\nu}$ pairs. We
employ the same number of integration points on both axes and split both
integration domains at the same energy, so that double integrals can be evaluated as
\begin{equation}
  \begin{split}
    \int & \dd k \int \, \dd k' \Hat{G}(k,k') =
      t^2 \sum_{j=1}^n \sum_{k=1}^n w_j w_k\times\\
      \bigg[
    &\Hat{G}\left(t \, x_j, t \, x_k\right) +
      \frac{\Hat{G}\left(t \, x_j, t / x_k\right)}{x_k^2} + \\
    +&\frac{\Hat{G}\left(t / x_j, t \, x_k\right)}{x_j^2} + 
       \frac{\Hat{G}\left(t / x_j, t / x_k\right)}{x_j^2 x_k^2}
       \bigg] \, .
  \end{split}
  \label{eq:gauss_legendre_integration}
\end{equation}
In the case of $\ep \, \em$ annihilations and NN bremsstrahlung, we set $t$ to half of the average energy of a
neutrino-antineutrino couple emitted by NN bremsstrahlung, namely, $t =
2.182\,T$ \cite{Burrows:2004vq}.
We use this value also when evaluating the contribution of such reactions to spectral
emissivities and inverse mean free paths via 1D integrals (see \secref{boltzmann_equation}).

Using the same quadrature prescription for different reactions allows us to sum together the
integrands associated to different processes
beforehand, avoiding to perform a specific integration for each of them.
In this respect, we checked that the
contribution of (inverse) $\ep \, \em$ annihilations
to the gray emissivities and opacities
converges quickly enough with respect to $n$ even if $t$ is not specifically
tailored to the process. Notice that the number of evaluations of $\Hat{G}$
in \eqref{gauss_legendre_integration} is equal to $4n^2$. However, 
since the quadrature nodes and
weights are the same along the two axes, we can exploit symmetry relations between
kernels with $(k,k')$ and $(k',k)$ dependence. This reduces by half the number
of kernel evaluations necessary, leading to a sizable speed up of the
execution.

In the case of \ac{NEPS} integration, setting $t$
to the above-mentioned value does not allow to achieve a quick convergence
with respect to $n$, since the distribution of the quadrature nodes across the
domains does not sample efficiently the integrands. In fact, the position where the
integrands for \ac{NEPS} peak depends on the energies of the neutrinos themselves.
Therefore, for the evaluation of spectral emissivities and inverse mean free paths
for \ac{NEPS} we set $t=k$, corresponding to the position where the value of the
integrands is the largest. For the computation of the gray quantities instead,
we follow a different paradigm than \eqref{gauss_legendre_integration}.
We first apply a rotation of $45^{\circ}$ on the integration variables,
\ie, $u = k + k'$ and $v = k - k'$, so that the double integrals can be rewritten as
\begin{equation}
    \begin{split}
        \int \dd k & \int \, \dd k' \Hat{G}(k,k') = \\
        & \frac{1}{2}  \int_{-\infty}^{+\infty} \dd u \int_{-u}^{+u} \dd v \,
	    \Hat{G} \left(\frac{u+v}{2},\frac{u-v}{2}\right) \, . 
    \end{split}
\end{equation}
We then split the innermost integral at $v=0$, rescaling the integration
variable into the $[0,1]$ interval for both segments and using $n'$
quadrature points along each of the two. In this way, the sampling of the $\Hat{G}$ function
is more efficient, since we are splitting the quadrature nodes evenly
across the position of the peak of the integrands, \ie, $v=k-k'=0$. Please
note that $n'$ can be in principle different than $n$, allowing us to use
a different number of integration points depending on the reaction. However,
in the context of this manuscript we always set $n'=n$.
We finally exploit \eqref{1d_gauss_legendre_integration} using
$t'{\sim} 8 T$ and $n'$ points to deal with
the outermost integration over the $[0,+\infty)$ interval, where $t'$
was fine-tuned to achieve the quickest convergence. In the end, the
formula implemented for evaluating the \ac{NEPS} contribution to gray emissivities
and opacities reads
\begin{equation}
    \begin{split}
         \int \dd k & \int \, \dd k' \Hat{G}(k,k') = 
         \frac{(t')^2}{2} \sum_{j=1}^{n'} \sum_{k=1}^{n'} w_j w_k\times\\
        \bigg[
        &x_j \Hat{G}\left(a_{jk},b_{jk}\right) +
        x_j \Hat{G}\left(b_{jk},a_{jk}\right) + \\
        &\frac{\Hat{G}\left(a_{jk}/x_k^2,b_{jk}/x_k^2\right)}{x_j^3} +
        \frac{\Hat{G}\left(b_{jk}/x_k^2,a_{jk}/x_k^2\right)}{x_j^3} \bigg] \, , 
    \end{split}
	\label{eq:NEPS_integration}
\end{equation}
where we defined $a_{jk}\equiv t' x_j(1-x_k)/2$ and $b_{jk}\equiv t' x_j(1+x_k)/2$ to
lighten the notation.

\begin{figure}
    \centering
    \includegraphics[width=\columnwidth]{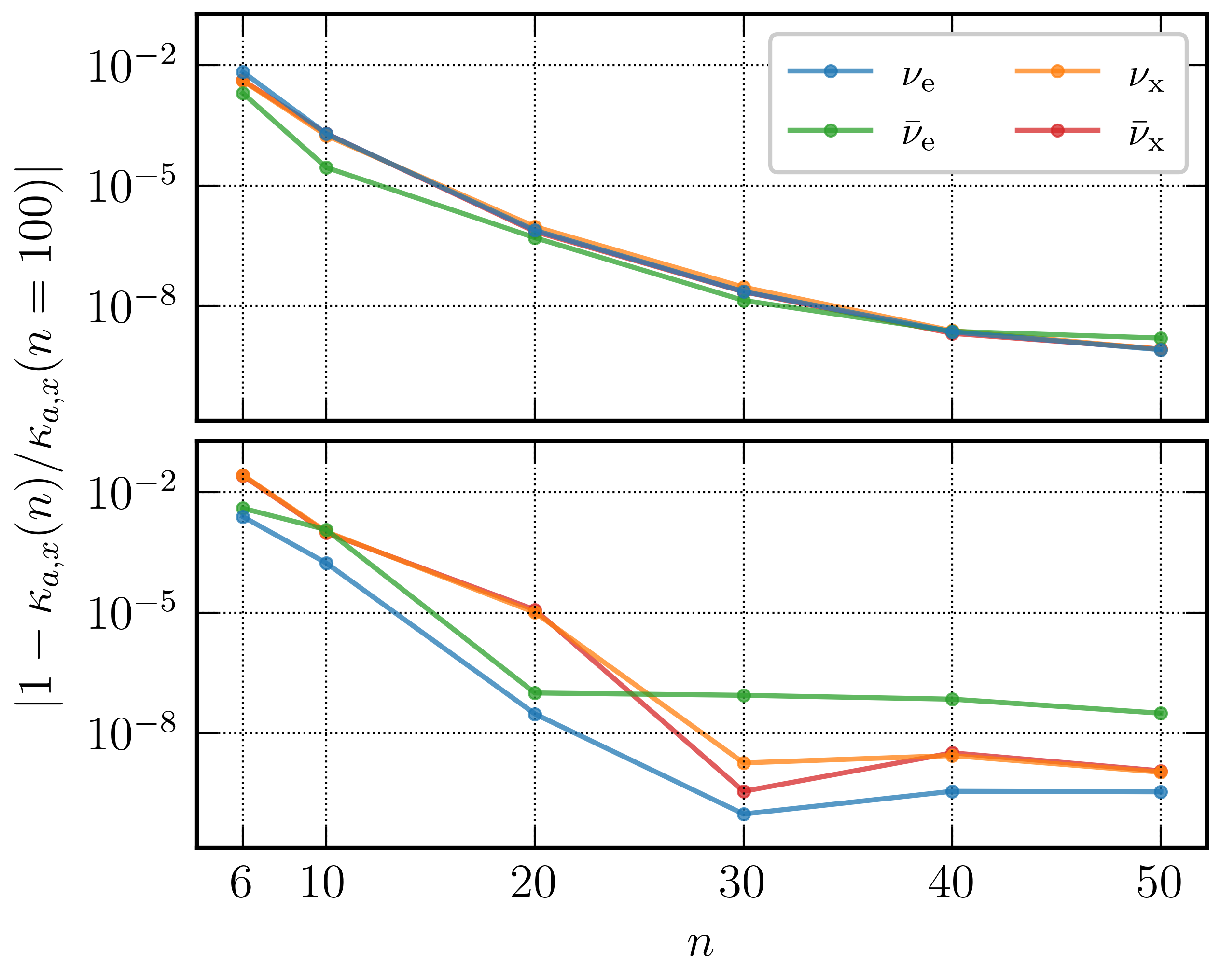}
    \caption{Top (bottom) panel: convergence level of the total $\kappa_{a,x}$ with
    respect to the number of quadrature points for the thermodynamic conditions
    at point A (F) in \tabref{thermodynamics_points}, using $n=100$ as a reference. Different colors refer to the
    different neutrino flavors.}
    \label{fig:quadrature_convergence}
\end{figure}

Figure~\ref{fig:quadrature_convergence} shows the convergence
in the calculation of the total gray absorption opacity with respect
to $n$, for the thermodynamic conditions at points A and F in
\tabref{thermodynamics_points} and reconstructing the distribution
function through \mone quantities. The relative error in the evaluation of $\kappa_{a}$ at
point A is below $10^{-2}$ for all neutrino species already when considering
$n=6$, which we identify as a good trade-off in terms of accuracy and performances
for the coupling in simulations of \bnsnurates to a gray \mone scheme
(see \secref{performance}). The largest discrepancy with respect to the
$n=100$ calculation is observed for heavy-type (anti)neutrinos at point F,
which is around ${\sim}2\%$ for $n=6$. This is because NEPS is 
the dominant reaction channel in such conditions and, differently than for the other flavors,
the $\nux$ mean energy does not match the energy at which the integral is split
($\langle E_{\nux} \rangle {\sim}2t'$, see \tabref{thermodynamics_points}), affecting
the resolution in the integrand sampling.
However, this order of convergence is still reasonable considering the level of accuracy achieved in neutrino
transport with \mone schemes. Moreover, the magnitude of the gray opacity at point
F is much lower given the smaller density, and so is the absolute error.
The level of convergence observed for the other
gray quantities, \ie, $\tilde{\eta}_x$, $\eta_x$, and $\tilde{\kappa}_{a,x}$,
is qualitatively similar to the one shown in \figref{quadrature_convergence}.

\section{OPTICAL DEPTH AND NEUTRINO SURFACE COMPUTATION}
\label{sec:tau_nusphere}

In order to compute the optical depths necessary to locate
neutrino surfaces for the neutrino species $x$, we rely on a variant of the classical Dijkstra's algorithm
\cite{Dijkstra1959} applied to the three-dimensional data extracted from the \ac{BNS} simulation
under consideration. This iterative algorithm stores at each iteration an estimate
of the optical depth, 
$\tau_{\tn{est},x}$, for each cell in the computational domain.
This information is augmented by a priority queue which stores the spatial
coordinates and the current value of $\tau_{\tn{est},x}$ for a variable number of cells.
The queue is sorted in order of ascending optical depth. Initially, every cell
is assigned infinite $\tau_{\tn{est},x}$, except for the boundary cells for
which $\tau_{\tn{est},x}=0$. The priority queue initially contains the coordinates
and the estimated optical depths of the boundary cells only. At every iteration, we pop the first
element of the queue, whose coordinates we denote as $\boldsymbol{x}$. Based on its
location, we compute a new optical depth's estimate, $\tau'_{\tn{est},x}$, for every first neighbor
of the corresponding cell:
\begin{equation}
  \tau'_{\tn{est},x}(\boldsymbol{x}_{n}) = \tau_{\tn{est},x}(\boldsymbol{x}) + \lambda_x^{-1}
  \sqrt{\gamma_{ij}\Delta\boldsymbol{x}^i \Delta\boldsymbol{x}^j}\,,
    \label{eq:opt_depth_estimate}
\end{equation}
where $\boldsymbol{x}_{n}$ is the position of the 
$n$th neighbor cell; $\gamma$ is
the spatial three-metric; $\Delta\boldsymbol{x} := \boldsymbol{x}_{n} -
\boldsymbol{x}$ is the coordinate distance between the two cells; and all
quantities on the right-hand side are evaluated at
$\boldsymbol{x}$. Neutrino reactions enter in
\eqref{opt_depth_estimate} through the quantity $\lambda_x^{-1}$, defined in \secref{neutrino_surface}.
We then compare $\tau'_{\tn{est},x}(\boldsymbol{x}_{n})$
with the previous estimate at the same cell.
If $\tau'_{\tn{est},x}(\boldsymbol{x}_{n}) < \tau_{\tn{est},x}(\boldsymbol{x}_{n})$, we substitute the latter with the
former and we add an element to the priority queue, storing the coordinates
$\boldsymbol{x}_{n}$ and the corresponding new optical depth's estimate. These steps are repeated until the queue contains no elements. The
algorithm is guaranteed to terminate because the optical depth is always
non-negative, therefore it always exhibits a minimum. As such, this procedure
allows for the calculation of the optical depth for each cell in the entire computational domain.

\section{ENERGY-DEPENDENT EMISSIVITIES AND INVERSE MEAN FREE PATHS FOR ADDITIONAL CONDITIONS}
\label{sec:additional_plots}

\begin{figure*}
  \centering
  \includegraphics[width=\textwidth]{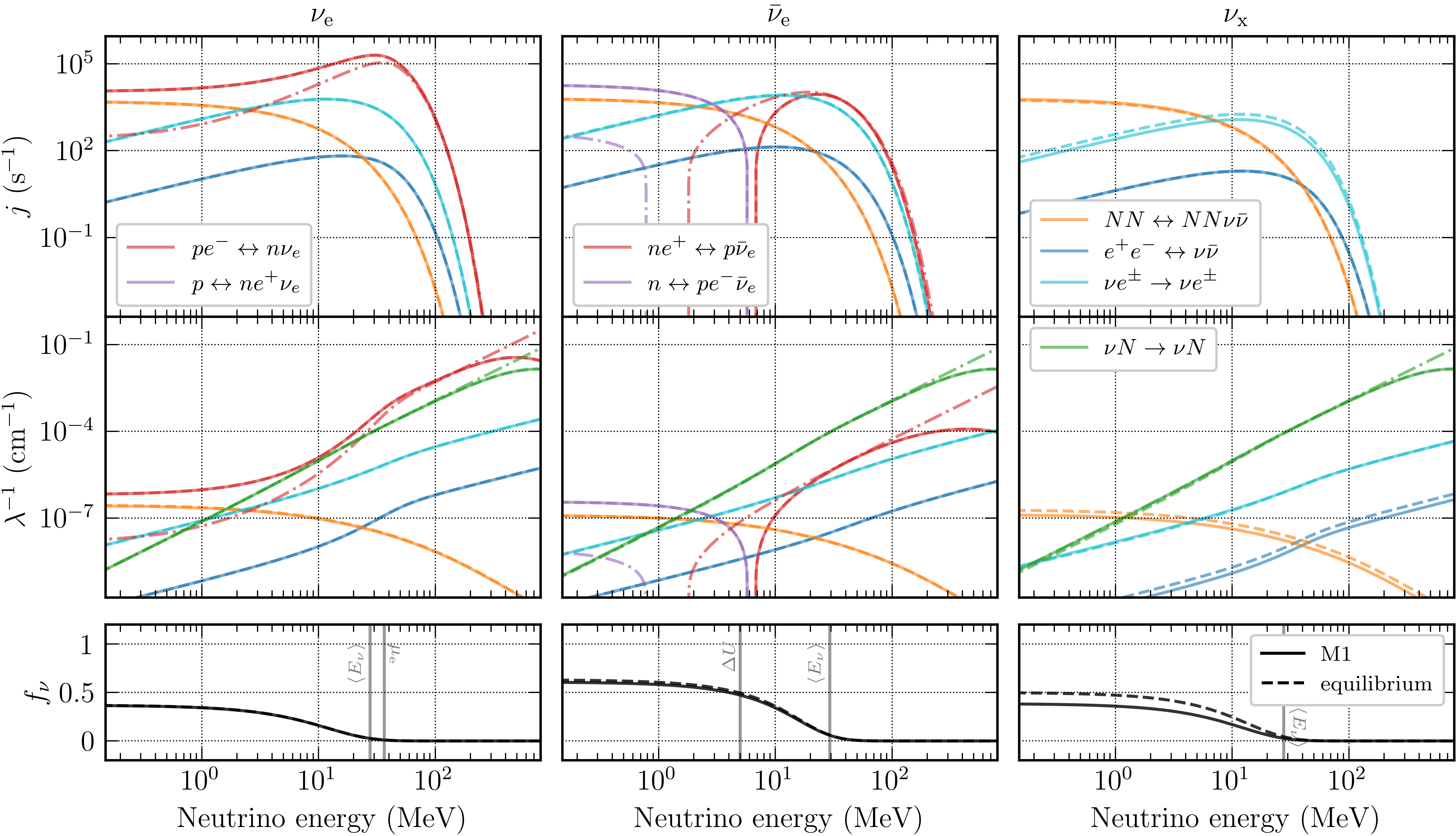}
  \caption{Same as
    \figref{DD2_M12980-12980_M1_LR_t_12390.4_n_50_rho_7.0e+14}
    but for thermodynamic point C in \tabref{thermodynamics_points}.}
  \label{fig:DD2_M12980-12980_M1_LR_t_12390.4_n_50_rho_1.0e+13}
\end{figure*}

\begin{figure*}
  \centering
  \includegraphics[width=\textwidth]{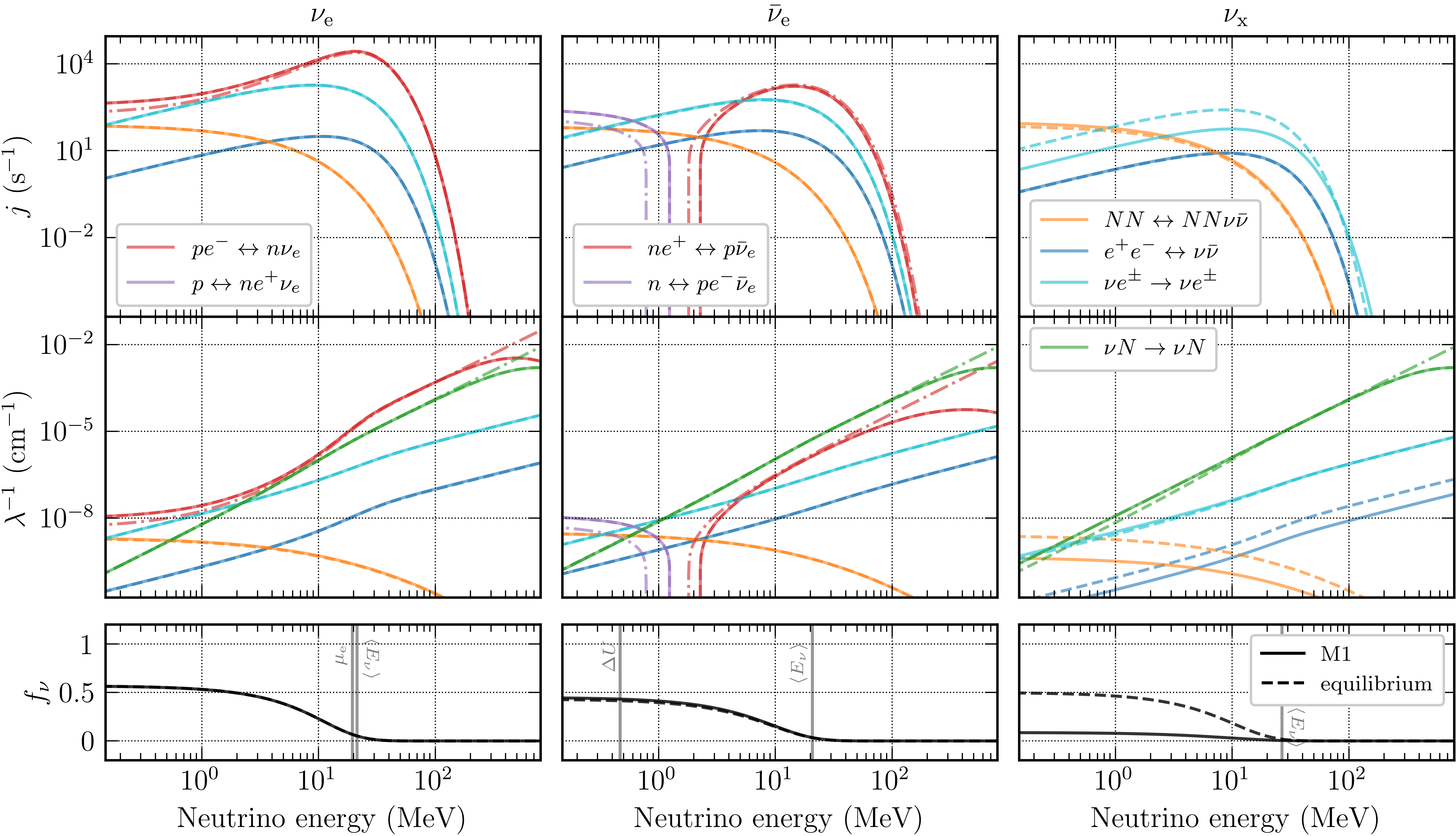}
  \caption{Same as
    \figref{DD2_M12980-12980_M1_LR_t_12390.4_n_50_rho_7.0e+14}
    but for thermodynamic point D in \tabref{thermodynamics_points}.}
  \label{fig:DD2_M12980-12980_M1_LR_t_12390.4_n_50_rho_1.0e+12}
\end{figure*}

\begin{figure*}
  \centering
  \includegraphics[width=\textwidth]{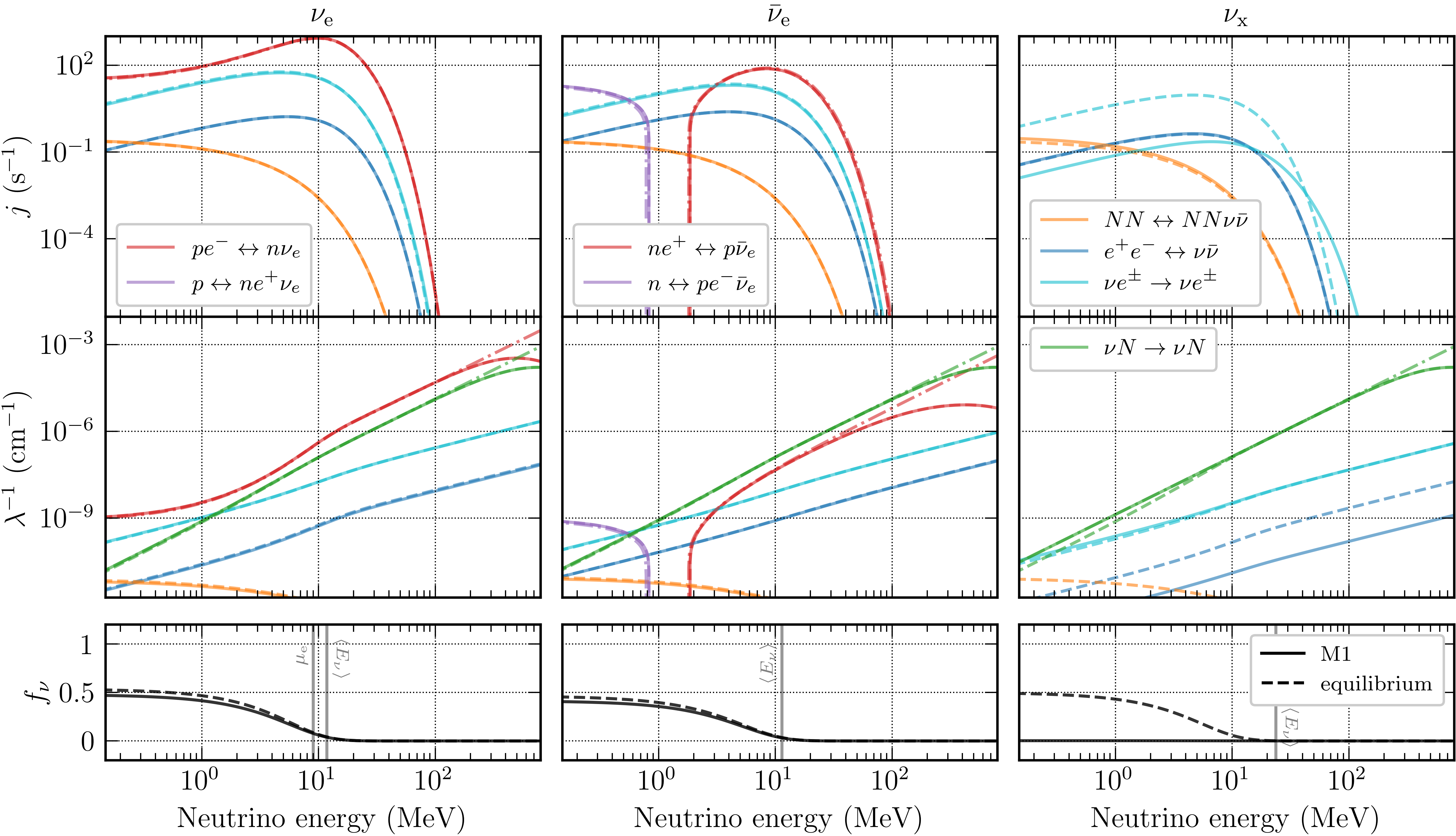}
  \caption{Same as
    \figref{DD2_M12980-12980_M1_LR_t_12390.4_n_50_rho_7.0e+14}
    but for thermodynamic point E in \tabref{thermodynamics_points}.}
  \label{fig:DD2_M12980-12980_M1_LR_t_12390.4_n_50_rho_1.0e+11}
\end{figure*}

We include in the following the plots exhibiting the spectral emissivities and
inverse mean free paths, as well as the neutrino distribution functions, for
thermodynamic conditions extracted at the points C, D, and E of \tabref{thermodynamics_points}
(Figs.~\ref{fig:DD2_M12980-12980_M1_LR_t_12390.4_n_50_rho_1.0e+13}-
\ref{fig:DD2_M12980-12980_M1_LR_t_12390.4_n_50_rho_1.0e+11},
respectively).
They complete the discussion in \secref{reaction_comparison} by showing the behavior of the various reactions
in the transition between optically thick and optically thin regimes.

\providecommand{\noopsort}[1]{}\providecommand{\singleletter}[1]{#1}%

\end{document}